\documentclass[12pt]{article}

\usepackage{amsfonts, amssymb, amsthm}
\usepackage[textures]{epsfig}

\newcommand{\sect}[1]{\setcounter{equation}{0}\section{#1}}
\newcommand{\subsect}[1]{\subsection{#1}}

\renewcommand{\theequation}{\arabic{section}.\arabic{equation}}

\newtheorem{tw}{Theorem}

\newtheorem{de}{Definition}

\newtheorem{co}{Corollary}

\newcommand{\be}{\begin{equation}}

\newcommand{\ee}{\end{equation}}

\newcommand{\bea}{\begin{eqnarray}}

\newcommand{\eea}{\end{eqnarray}}

\begin{document}

\title{ QUANTIZATION ON
 A $2$--DIMENSIONAL
PHASE SPACE WITH  A  CONSTANT CURVATURE TENSOR}
\author{M. Gadella, M. A. del Olmo  and J. Tosiek~\footnote{On leave of
absent from Tech. Univ. of  \L\'{o}d\'{z}, Poland} \\ {\em  Departamento de  
F\'{\i}sica Te\'orica}\\
 {\em Universidad de  Valladolid}\\
 {\em E-47011, Valladolid,  Spain}\\[0.5cm]
{\rm E.mails: gadella@fta.uva.es, olmo@fta.uva.es, tosiek@ck-sg.p.lodz.pl}}
\date{}
\maketitle
\begin{abstract}
Some properties of the star product of the Weyl type (i.e.
associated with the Weyl ordering)  are proved. Fedosov
construction of the $*-$product on a $ 2$--dimensional phase space
with a constant curvature tensor is presented.
 Eigenvalue equations  for  momentum $ p $ and position $q$ on a
$ 2$-dimensional phase space with  constant  curvature tensors
  are solved.
\end{abstract}


\sect{Introduction}

\setcounter{equation}{0}

The  traditional formulation of quantum mechanics, in which the
space of pure states is a complex Hilbert space ${\cal H}$ and
observables are represented by self-adjoint operators acting on
${\cal H},$ is mainly applied  for systems, whose classical phase
spaces are $ \mathbb{R}^{2n}. $ In other cases the Dirac
quantization programme \cite{dir} is out of work \cite{jos},
\cite{che}. To deal with quantum description of systems with
nontrivial phase spaces it is necessary to use other methods:
geometric quantization \cite{van}, \cite{woo}, quantization by
Stratonovich--Weyl kernels \cite{stra}--\cite{jarema} or
 deformation quantization.
 We shall apply the last formalism.

In the deformation quantization approach, observables and states
are represented by functions or generalized functions on the
cotangent manifold $T^*{\cal M},$ where ${\cal M}$ is a given
manifold called the  base space. The space $T^*{\cal M}$ plays
here the role of phase space of the system under our
consideration.

The basic idea in the deformation quantization method is the
replacement of the standard commutative product of functions by a
noncommutative product, usually called the star product and
denoted by $*.$ This replacement is called `deformation' and the
$*$--product (or  deformed product) depends on a parameter,
usually the Planck constant $\hbar.$

Information about a state of a system is represented by some
functional over a space of observables. In case of the phase space
$ \mathbb{R}^{2n} $ such a functional is called {\it the Wigner
function}, so we shall use this term although  the functional
representing  a quantum state in fact need not be a function on $
T^*{\cal M}, $ but it is often a generalized function or
distribution.

The formulation of  quantum mechanics on  phase space was first
proposed by   Wigner in the 30's \cite{1q} and later by Moyal
\cite{moy} but the progress of this method was fostered in 70's by
Agarwal, Wolf \cite{caj} and  by Bayen {\em et al} \cite{cak}.
Since that time several monographs have been written on this
subject \cite{car}--\cite{lee}.

In the case of  the flat phase space $\mathbb{R}^{2n}$ the
$*$--product was defined by Groenewald \cite{gro} and Moyal
\cite{moy}. This is the reason, why for systems with  phase space
$\mathbb{R}^{2n}$ the $*-$ product is also known as {\it the Moyal
product}.

In this work, we study physical systems with curved phase spaces.
To do it we use the Fedosov  procedure  \cite{4,5} of the
construction of the $ *-$ product for any symplectic space. This
approach to the deformation quantization has been propagated by
many  authors \cite{bo1}--\cite{ca2}. Another methods of
construction of quantum mechanics on Poisson manifolds can be
found  in \cite{ki1}--\cite{tam}.

For the same physical system there exist many $*$--products (see
\cite{caj,3}). Such a multitude is the consequence of the choice
of ordering \cite{trz,2}. In this paper we consider the Weyl
ordering. The $*$--product for this ordering is called
$*$--product of the Weyl type but we will simply call it
$*$--product.

  In classical mechanics  we have two fundamental structures: a
  symplectic manifold that plays the role of phase space and a
  Hamilton function on this manifold.
In quantum mechanics a new element appears:  curvature of the
phase space.  This curvature affects  the definition of a
$*$--product. It means that starting from the same classical phase
space we are able to build different quantum systems.

There is not a unique way to construct a geometric structure on $
T^*{\cal M}.$ We have chosen our construction based on the
following assumptions:
\begin{enumerate}
\item
\label{zal11} we equip the phase space $ T^*{\cal M}$  with some
symplectic structure $ \omega $ only. We do not define any metric
structure on it;
\item
\label{zal12} the Riemannian structure defined on the base space
${\cal M} $ affects  the symplectic connection on $ T^*{\cal M}.$
The precise definition of the influence mentioned above is
presented in section \ref{abe}.
\end{enumerate}

This paper consists of two parts. The first one includes some
basic information and properties necessary to deal with a problem
solved in the second part. We start with the  foundations of the
Fedosov formalism. We define the Weyl algebra bundle $ {\cal V} $,
 construct the abelian connection in $ {\cal V} $ and
build the $ * $--product of observables. We introduce the trace
density and  the theorem  (\ref{teorema}), which give us the
opportunity to compute this density in some cases. We do not
present the full mathematical structure of Fedosov idea but only
the basic ideas that will allow us to follow the computations done
in next paragraphs. Section \ref{drugi} contains some properties
of the $*$--product necessary for our further consideration.

The main part of our work consists in the application of the
Fedosov formalism to a  $2$--dimensional ($2$--D) phase space with
globally constant curvature tensor. This is the most elementary
generalization of the phase space $ \mathbb{R}^2$ under conditions
(\ref{zal11}, \ref{zal12}). We consider the case of constant
curvature tensor instead of the constant scalar curvature  because
for every symplectic manifold the   scalar curvature vanishes.

 We shall look for the
eigenvalues and the Wigner eigenfunctions of the momentum and the
position on such a phase space. To find the eigenvalue equations
of the momentum and the position, we follow the route proposed by
Fedosov.
 We start from the construction of some symplectic connection
 $ \nabla$ compatible with assumptions (\ref{zal11}, \ref{zal12})
 on $ T^*\mathbb{R}. $
Having this connection, we build an abelian connection in the Weyl
algebra bundle $ {\cal V}. $ Finally using the abelian connection,
we present and solve  the eigenvalue equation for  momentum $ p. $
This equation is a differential equation in an infinite range,
which cannot be easily solved. For this reason   we substitute it
by the eigenvalue equation for $ p^2$, which   is similar to the
modified Bessel equation. We  find its solution and prove that the
obtained result  is really the eigenfunction of the  momentum. By
analogy, we write the   eigenvalue equations for  momentum $ p $
and position $ q $ for a phase space with constant positive
curvature tensor. We compare our results with the solutions for
the flat symplectic space $ \mathbb{R}^2 $ and show that
eigenfunctions of the momentum and position in the flat case are
limits of the solutions obtained  for the curved phase spaces
analyzed by us. We also explain, why the presented considerations
cannot be reduced to the problem of the eigenvalues of momentum
and the position on $\mathbb{R}^2,$ although all the $*$--products
on every $2$--D symplectic space are equivalent. We present also
the eigenvalue equation and its solution for the position.
Some  computations are rather long so we decided to present them
in Appendices.

The formulas appearing in the Fedosov definition of star product
are very pleasant for numerical analysis, hence some of the
results presented in this contribution have been computed by  using
 `Mathematica 4.0'. There is a more completed  version with eight figures 
that the interested
reader can  request to the authors


\sect{The $ * $--product on curved phase spaces}

\label{drugi}

This section contains the basic information necessary for later
use. It is presented in  the form of a set of useful formulas
rather than a rigorous presentation of mathematical aspects of
construction of the star product.
 The reader interested in  details and proofs should see \cite{4} or \cite{5}.

Let $ T^*{\cal M} $ be a $ 2n$--D cotangent bundle over
a manifold $ {\cal M}. $ The manifold $ T^*{\cal M} $ is equipped
with some symplectic structure $ \omega, $ which in local
coordinates is written as 

\be \label{symp} \omega =
\frac{1}{2}\omega_{ij} dq^i \wedge dq^j, \qquad  i,j=1,\ldots, 2n.
\ee

Assume that $ T^*{\cal M} $ is the phase space of
some physical system.

\begin{de}
\label{de00} An observable $ F(q^i), \; i=1, \ldots,2n, $ is a
real smooth function on   $ T^*{\cal M}. $
\end{de}

We do not distinguish between classical and quantum observables.
Both of them are defined in the same way.

Let us start constructing the $*$--product on the symplectic
manifold $ T^*{\cal M}. $ For  every point $ {\tt p} \in T^*{\cal
M}$, we define some formal series with respect to a parameter
denoted by $ \hbar $ 
\be \label{12} a \; := \; \sum^{\infty}_{l=0}
\hbar^k a_{k,i_1 \ldots i_l} X^{i_1}_{\tt p} 
\cdots X^{i_l}_{\tt p}, \qquad k \geq 0. 
\ee 
For $ l=0, $ we put $ a= \hbar^k a_k. $

The symbols in the formula (\ref{12}) mean:
$ \hbar $ is a positive parameter, that  we identify with  the
Planck constant;
$ X^1_{\tt p}, \ldots, X^{2n}_{\tt p} $ are the components of an
arbitrary fixed vector belonging to the tangent space $ T_{\tt
p}(T^*{\cal M}) $ to the symplectic manifold $ T^*{\cal M} $ at
the point $ {\tt p}. $ The components $ X^1_{\tt p}, \ldots,
X^{2n}_{\tt p} $ have been written in the natural basis $ \left(
\frac{\partial}{\partial q^i}\right)_{\tt p} $ determined by the
chart $ (W_{\lambda}, \varrho_{\lambda}), $ such that $ {\tt p}
\in W_{\lambda} $ ;
$ a_{k,i_1 \ldots i_l} $ are the components of some covariant
tensor symmetric with respect to indices $ (i_1, \ldots, i_l) $ in
the basis $ dq^{i_1} \otimes  \cdots \otimes dq^{i_l}. $

Let $ V(X_{\tt p}) $ be the set of all elements of  kind
(\ref{12}) at the point $ {\tt p} \in T^*{\cal M}. $ Then, the
tetrad $ (V(X_{\tt p}), \mathbb{C},+,\cdot) $ is a linear space.
In  $ V(X_{\tt p})$ we define a new product `$\circ$': \be
\label{13} a \circ b  \; :=\sum_{t=0}^{\infty} \left( \frac{i
\hbar}{2}\right)^t \frac{1}{t!} \omega^{i_1 j_1} \cdots
\omega^{i_t j_t} \frac{\partial^t a}{\partial X^{i_1}_{\tt p}
\ldots \partial X^{i_t}_{\tt p}} \frac{\partial^t b}{\partial
X^{j_1}_{\tt p} \ldots \partial X^{j_t}_{\tt p}}. \ee The tensor $
\omega^{ij} $ fulfills the equation \be \omega^{i j} \omega_{jk}
= \delta^i_k, \ee where $\omega_{jk} $ are the coefficients of the
symplectic form (\ref{symp}) in the basis $ dq^{i} \otimes
dq^{j}. $
The definition of the $ \circ$--product  is chart independent.
\begin{tw}
\label{tw00} The pair  $ (V(X_{\tt p}),  \circ) $ is  an
associative noncommutative algebra with  unit.
\end{tw}
We will denote it by $ {\cal V}( X_{\tt p}) $ and call it {\it the
Weyl algebra}. The unit in $ {\cal V}( X_{\tt p}) $ is just a
formal series equal to $1. $

Let us consider the set \be \label{14} {\cal V} \; := \;
\bigcup_{{\tt p} \in T^*{\cal M}} {\cal V}( X_{\tt p}). \ee
The new object ${\cal V}$ has the structure of the vector bundle.
Its base space is the manifold $ T^*{\cal M}$ and the fibre is
isomorphic to the vector space $ (V(X_{\tt p}),
\mathbb{C},+,\cdot). $ Moreover the fibre is also an algebra. The
bundle ${\cal V}$ is known as {\it the Weyl bundle}.
We are able to define differential forms with values in ${\cal V}.$

\begin{de}
The $m$--differential form with values in the Weyl bundle $ {\cal
V}$ is given by the expression \be \label{15} a =
\sum^{\infty}_{l=0} \hbar^k a_{k,i_1 \ldots i_l,j_1 \ldots
j_m}(q^1, \ldots, q^{2n}) X^{i_1}
 \cdots X^{i_l}
dq^{j_1} \wedge \cdots \wedge dq^{j_m},
\ee
where $ 0 \leq m \leq 2n. $
\end{de}
Here, $ a_{k,i_1 \ldots i_l,j_1 \ldots j_m}(q^1, \ldots, q^{2n}) $
is a smooth tensor field symmetric with respect to indices $ (i_1,
\ldots, i_l) $ and antisymmetric with respect to $ (j_1, \ldots
,j_m). $ To simplify the notation we will omit  the coordinates
$(q^1, \ldots, q^{2n}).$

The product of forms like (\ref{15}) is  the multiplication
(\ref{12})  of elements of the algebra $ {\cal V} $ and the
external product $ `\wedge $' of forms.
\begin{de}

The commutator of  the forms $ a \in {\cal V} \otimes
\Lambda^{m_1}$ and $ b \in {\cal V} \otimes \Lambda^{m_2}$ is a
form $ [a,b] \in {\cal V} \otimes \Lambda^{m_1 + m_2}$ defined by
\be \label{16} [a,b] \; := \; a \circ b -(-1)^{m_1 \cdot m_2} b
\circ a. \ee
\end{de}

The form $ a \in {\cal V} \otimes \Lambda^{m} $ is called {\it
central}, if for every $ b \in {\cal V} \otimes \Lambda $ the
commutator $ [a,b] $ vanishes. It is possible to prove that only
forms which do not contain elements $ X^i $ are central.

To construct the $* $--product on $ T^*{\cal M} $ requires 
 a new operator $ \delta^* : \; {\cal V} \otimes \Lambda^{m}
\rightarrow {\cal V} \otimes \Lambda^{m-1} . $ The full
explanation of this step requires representation of $ a \in {\cal
V} \otimes \Lambda^{m} $ in a form similar to the Hodge--De Rham
decomposition, so we shall present only final formulas.

\begin{de}
The operator $ \delta^* \; : \; {\cal V} \otimes \Lambda^{m}
\rightarrow {\cal V} \otimes \Lambda^{m-1} $ fulfills the relation
\be \label{17} \delta^* a := X^k \left( \frac{\partial}{\partial
q^k} \right) \rfloor a. \ee
\end{de}

Having  $ \delta^* ,$  we  define a new operator $\delta^{-1}$ as
\be \label{delta-1}
 \delta^{-1} := \left\{ \begin{array}{cc}
 \frac{\delta^*}{m+l} &  {\rm for }\;\; l+m>0 \\[0.3cm]
 0  &{\rm for }\;\; l+m=0. \end{array} \right. \ee By
 $ l $ we denote  the
degree of expression in $ X^i $ and by $ m $  the degree of the
form. The sum $l + m$ is the range of the tensor $a_{k,i_1 \ldots
i_l,j_1 \ldots j_m}.$

\vspace{0.5cm}
 Let us define
the parallel transport on $ T^*{\cal M}. $
\begin{de}
The symplectic connection $ \nabla $ on the phase space $ T^*{\cal M} $ is a torsion-free connection satisfying the condition
\be
\label{-18}
\nabla_{X}\omega=0,
\ee
where $ X \in T(T^*{\cal M}) $ is a tangent vector to an arbitrary curve along which the form $ \omega $ is parallel transported.
\end{de}
In a local chart, the expression (\ref{-18}) has the form \be
\label{18} \omega_{ij;k}=0, \ee where the semicolon `$ ; $' stands
for the covariant derivative.

In Darboux coordinates  the formula (\ref{18}) is equivalent to
the system of equations \be \label{19} \omega_{ij;k}=
-\Gamma^l_{ik} \omega_{lj} -\Gamma^l_{jk} \omega_{il}
=\Gamma_{jik}-\Gamma_{ijk}=0. \ee We define \be \label{20}
\Gamma_{ijk} := \omega_{il} \Gamma^l_{jk}. \ee In  Darboux
coordinates  the coefficients $ \Gamma_{ijk} $ are symmetric with
respect to indices $ (i,j,k). $

\begin{tw}
\label{tw02} A symplectic connection is not unique (
\cite{4},\cite{5},\cite{3}). In  local coordinates the difference
between two symplectic connections $\Gamma_{ijk} $ and $
\tilde{\Gamma}_{ijk}$ \be \Delta_{ijk} := \Gamma_{ijk} -
\tilde{\Gamma}_{ijk} \ee is a tensor symmetric with respect to
indices $(i,j,k).$
\end{tw}

\begin{de}
The exterior covariant derivative $
\partial: {\cal V} \otimes \Lambda^{m} \longrightarrow
{\cal V} \otimes \Lambda^{m+1} $ is a linear operator defined by
the formula \be \label{22}
\partial a := dq^r \wedge a_{;r}, \qquad a \in {\cal V} \otimes \Lambda^{m}.
\ee
\end{de}

In the original Fedosov paper \cite{4}   the operator $\partial$ is
called `the connection on the bundle $ {\cal V}$'.

Let $ d $ be the exterior derivative and let $ \Gamma $ be the $
1- $ form
  of connection  equal to
\be \label{23} \Gamma =\frac{1}{2} \Gamma_{ijk} X^i X^j dq^k . \ee

\begin{tw}
\label{tw03}
In a Darboux atlas, the exterior covariant derivative is given by
\be
\label{24}
\partial a= da + \frac{1}{i \hbar}[\Gamma,a].
\ee
\end{tw}

The curvature of the connection $ \Gamma $ is defined as the $ 2-$
form \be \label{25} R = d \Gamma + \frac{1}{2i \hbar}
[\Gamma,\Gamma]. \ee In local coordinates \be \label{26} R =
\frac{1}{4} \omega_{im} R^m_{jkl}X^i X^j dq^k \wedge dq^l, \ee
where \be \label{27} R^m_{jkl} := \frac{\partial
\Gamma^m_{jl}}{\partial q^k} -\frac{\partial
\Gamma^m_{jk}}{\partial q^l} +\Gamma^u_{jl}\Gamma^m_{u k} -
\Gamma^u_{jk} \Gamma^m_{ul}. \ee The coefficients $ \Gamma^i_{jk}
$ are determined by the elements $ \Gamma_{ijk} $ via relation
(\ref{20}).

It is possible to show that for every form   $  a \in {\cal V}
\otimes \Lambda^{m}, $ the second exterior covariant derivative is
given by \be \label{28}
\partial (\partial a)= \frac{1}{i \hbar} [R,a].
\ee

To construct the Moyal product we need to create some new `flat'
connection
 on $ T^*{\cal M}.$

\begin{de}
\label{de101} The connection $\tilde{\nabla}$  in the bundle $
{\cal V} $ is called abelian, if the exterior covariant derivative
$ D$ defined by the formula \be \label{29} Da := da + \frac{1}{i
\hbar}[\tilde{\Gamma},a], \ee fulfills the equation \be \label{30}
D(Da)=  \frac{1}{i \hbar}[\Omega,a]=0 \ee for every $ a \in {\cal
V} \otimes \Lambda. $ Here $\tilde{\Gamma}$ is the $1-$form
representing the abelian connection $\tilde{\nabla}$ in a Darboux
chart $(W_{\varrho}, \varphi_{\varrho})$ and    $ \Omega $ is the
curvature of the connection $ \tilde{\Gamma} $ \be \label{31}
\Omega :=
 d \tilde{\Gamma} +  \frac{1}{2 i \hbar}[\tilde{\Gamma},\tilde{\Gamma}].
\ee
\end{de}

Formally, any abelian connection is similar to a flat one because
in both cases
 \[
 a_{;sd}= a_{;ds} \quad{\rm for} \qquad
1 \leq s,d \leq 2n.
\]
 However the  abelian curvature can be a central form different
from $ 0. $

Let us assume that the exterior covariant derivative with an
 abelian connection $ \tilde{\Gamma} $ can be written
in the form

\be
\label{32}
Da= \partial a + \frac{1}{i \hbar}[r +\omega_{ij} X^i dq^j,a],
\ee
where $ r \in  {\cal V} \otimes \Lambda^1. $

One can show that if $ \delta^{-1}r=0, $ then $ r $ is defined by
the recurrent formula \be \label{33} r = \delta^{-1} R +
\delta^{-1} \left( \partial r + \frac{1}{i \hbar} r^2 \right). \ee
The first element in the  series $ r $ is the 1-form $ \delta^{-1}
R. $ The square $ r^2 $ is computed in the bundle $ {\cal V}  $,
so $ r^2= r \circ r.$

The following theorem holds.
\begin{tw}
The set of forms $ a \in {\cal V} \otimes \Lambda $ such that $
Da=0 $ is a subalgebra ${\cal V}^D \otimes \Lambda $, of the
algebra $ {\cal V} \otimes \Lambda. $
\end{tw}

Let $ \sigma(a) $ be a `projection' of a $ 0 $--form $ a \in
{\cal V} $ onto the phase space  $ T^*{\cal M} $, i.e. a mapping
 assigning to $ a $ its components that do not contain $ X^1,
\ldots, X^{2n}. $ The mapping $ \sigma $ is unique.

It is possible to construct a  relation between some subset of
smooth functions $ {\cal F}_S(T^*{\cal M})$ on
 $ T^*{\cal M} $ and $ 0 $--forms belonging to ${\cal V}^D. $
The mapping  from ${\cal V}^D $ onto $ {\cal F}_S(T^*{\cal M}) $
is just the projection $ \sigma $ to the inverse mapping
 we will denote
by $  {\sigma}^{-1}. $

We used the phrase `some subset of smooth functions' because for
functions containing negative powers of the deformation parameter
$ \hbar $
 the mapping $  {\sigma}^{-1}  $ may not exist.
From the other side, we do not want to restrict ourselves to
functions which are series in $ \hbar $ because Wigner functions
usually contain negative powers of $ \hbar. $

\begin{tw}
Every  smooth function  $ F \in {\cal F}_S(T^*{\cal M}) $
 defines uniquely the element $ \sigma^{-1}(F) $
of the bundle $ {\cal V}^D $
such that $ \sigma\left( \sigma^{-1}(F) \right)=F. $
\end{tw}

The element $ \sigma^{-1}(F) $ fulfills the recurrent relation \be
\label{34} \sigma^{-1}(F)= F + \delta^{-1} \left( \partial
(\sigma^{-1}(F)) + \frac{1}{i \hbar}[r,\sigma^{-1}(F)] \right).
\ee

Now we are able to construct the $ * $--product on $ T^*{\cal M}$.

\begin{de}
Let $ F_1, F_2 \in {\cal F}_S(T^*{\cal M}).$
 The star product  of $ F_1, F_2 $
 is defined by
\be \label{35} F_1 * F_2  := \sigma \left( \sigma^{-1}(F_1) \circ
\sigma^{-1}(F_2) \right). \ee
\end{de}

 Although the above definition and theorems presented in this section 
are valid for functions
 belonging to ${\cal F}_S(T^*{\cal M}), $ we will apply these results
 to generalized functions.

Let $ W(q^i) $ be the functional representing the state of a
quantum system. We  call it the Wigner function for this state.
The average value of an observable $ F(q^i) $ in $ W(q^i) $ is
defined as \be \label{wi1} \langle F \rangle = \int_{T^*{\cal M}}
W(q^i)
* F(q^i) t(q^i)dq, \ee where $dq= dq^1 \cdots dq^{2n}.$ The
integral (\ref{wi1}) represents the relation 
\be \label{wi2}
 \langle \hat{F} \rangle = {\rm Tr} (\hat{F} \cdot \hat{\varrho} ), 
\ee 
the operator
$\hat{\varrho}$ is the density operator. Of course the operation
`Tr' does not depend on the order of the operators, so that the
same property must be fulfilled by the integral (\ref{wi1}). The
function $ t(q^i) $ is responsible for the following fact:  for
any two functions $F_1(q^i), F_2(q^i),$ for which the integral \be
\int_{T^*{\cal M}} F_1(q^i)
* F_2(q^i) t(q^i)dq \ee is well defined, the following relation
holds \be \int_{T^*{\cal M}} \{ F_1(q^i) , F_2(q^i) \}_{M}
t(q^i)dq =0. \ee The symbol $ \{ F_1(q^i) , F_2(q^i) \}_{M}$
denotes the Moyal bracket 
\be
 \{ F_1(q^i) , F_2(q^i) \}_{M} := \frac{1}{i \hbar} \left( F_1(q^i) * F_2(q^i)
 - F_2(q^i) * F_1(q^i) \right).
\ee

\begin{de}
The function \be \label{tr2} t(q^i)= t_0(q^i) + \hbar t_1(q^i) +
\hbar^2 t_2(q^i) + \ldots  \;\; \in {C}^{\infty}(T^*{\cal M}) \ee is
called trace density.
\end{de}
The trace density is connected with an object known as the trace
of the Weyl algebra.
\begin{de}
Let $ {\cal V}^D_c \subset {\cal V}^D $ be an ideal consisting of
compactly supported flat sections. A linear functional on ${\cal
V}^D_c$ with values in formal Laurent series with negative powers
of $\hbar$ not exceeding $ n = \frac{1}{2} \dim T^*{\cal M}$ is
called a trace if it vanishes on commutators. Thus, 
\be
\label{tr1} {\rm Tr}  \: a = \sum_{k=0}^{\infty} \hbar^{k-n} c_k,
\ee where $c_k \in {\mathbb C}$ are constant coefficients and \be
{\rm Tr} \; (a \circ b) = {\rm Tr}  \; (b \circ a) \;\; {\rm for
}\;\; a,b \in {\cal V}^D_c. \ee
\end{de}
\begin{tw}
\label{teorema} For any smooth function $F \in {\cal F}_S(T^*{\cal
M}) $ such that  $Tr (\sigma^{-1}(F))$ exists the following
relation holds \be \label{tr3} {\rm Tr}(\sigma^{-1}(F)) = (2 \pi
\hbar )^{-n} \int_{T^*{\cal M}} F(q^i) t(q^i) dq. \ee
\end{tw}

 The construction of the trace density function in the general case is very complicated. 
The reader can find the
  full algorithm in \cite{5}. In our case, it is
sufficient to use the following statement:
\begin{tw}
\label{wazne}
 For every algebra ${\cal V}^D$ there exists a trace density 
(\ref{tr2}) fulfilling (\ref{tr3}). The coefficients $t_0(q^i),t_1(q^i), \ldots $ 
are polynomials
in the symplectic curvature tensor
$R_{ijkl}$ and its covariant derivatives.
\end{tw}


\sect{ Properties of the $*$--product}

\setcounter{equation}{0}

In this section, we present some general properties of the
$*$--product
 of functions on
$ T^*{\cal M} $  and of the  $ \circ $--multiplication  of forms.
We are not going to present a systematical review of these
concepts. We shall only mention  properties used in next
paragraphs. Other results can be found in \cite{bo0}.

\begin{tw}
\label{tw21} The commutator $[a,b] $ of  forms $a \in {\cal V}
\otimes \Lambda^{m}$ and $
 b \in {\cal V} \otimes \Lambda^{w}$ contains only terms with odd number
of derivatives in $ X^i. $
\end{tw}

\noindent
\underline{Proof}

The $  \circ$--product  is bilinear and, therefore, it is
sufficient to analyze the multiplication of  two terms
\[\begin{array}{l}
a=  a_{k,u_1 \ldots u_l,j_1 \ldots j_m}(q^1, \ldots, q^{2n}) X^{u_1}
 \cdots X^{u_l}
dq^{j_1} \wedge \cdots \wedge dq^{j_m}
\\[0.3cm]
b=  b_{z,v_1 \ldots v_d,r_1 \ldots r_w}(q^1, \ldots, q^{2n})
X^{v_1}
 \cdots X^{v_d}
dq^{r_1} \wedge \cdots \wedge dq^{r_w} .
\end{array}\]
To simplify the notation we will omit the variables $(q^1, \ldots, q^{2n}).$

From the definition (\ref{13}) of the $ \circ$--product
\be \label{d21}
\begin{array}{l}
a \circ b = \displaystyle
\sum_{t=0}^{\infty} \left( \frac{i \hbar}{2}\right)^t
\frac{1}{t!}
\omega^{i_1 p_1} \cdots \omega^{i_t p_t}
\frac{\partial^t (a_{k,u_1 \ldots u_l,j_1 \ldots j_m} X^{u_1}
 \cdots X^{u_l}
)}{\partial X^{i_1} \ldots \partial X^{i_t}}  
\\[0.3cm] 
\qquad \times \;  \displaystyle\frac{\partial^t (b_{z,v_1 \ldots v_d,r_1 \ldots
r_w} X^{v_1}
 \cdots X^{v_d}
)}{\partial X^{p_1} \ldots \partial X^{p_t}} (dq^{j_1} \wedge
\cdots \wedge dq^{j_m}) \wedge (dq^{r_1} \wedge \cdots \wedge
dq^{r_w}). 
\end{array}
\ee 
Analogously,
\be\label{d22}
\begin{array}{l}
b \circ a = \displaystyle
\sum_{t=0}^{\infty} \left( \frac{i \hbar}{2}\right)^t
\frac{1}{t!} \omega^{i_1 p_1} \cdots \omega^{i_t p_t}
\frac{\partial^t (b_{z,v_1 \ldots v_d,r_1 \ldots r_w} X^{v_1}
 \cdots X^{v_d}
)}{\partial X^{i_1} \ldots \partial X^{i_t}}  
\\[0.3cm]
\qquad \times \;  \displaystyle\frac{\partial^t (a_{k,u_1 \ldots u_l,j_1 \ldots
j_m} X^{u_1}
 \cdots X^{u_l}
)}{\partial X^{p_1} \ldots \partial X^{p_t}} (dq^{r_1} \wedge
\cdots \wedge dq^{r_w})
 \wedge
(dq^{j_1} \wedge \cdots \wedge dq^{j_m}) . 
\end{array}
\ee 
 We sum with respect
to the  indices $ i_1, \ldots , i_t , p_1 , \ldots , p_t, $ so
that we are able to exchange symbols $ i_g \leftrightarrow p_g $
for $ g=1, \ldots, t.$ Now 
\[\begin{array}{l}\displaystyle
\sum_{t=0}^{\infty} \left( \frac{i \hbar}{2}\right)^t \frac{1}{t!}
\omega^{i_1 p_1} \cdots \omega^{i_t p_t} \frac{\partial^t
(b_{z,v_1 \ldots v_d,r_1 \ldots r_w} X^{v_1}
 \cdots X^{v_d}
)}{\partial X^{i_1} \ldots \partial X^{i_t}} \cdot
\frac{\partial^t (a_{k,u_1 \ldots u_l,j_1 \ldots j_m} X^{u_1}
 \cdots X^{u_l}
)}{\partial X^{p_1} \ldots \partial X^{p_t}} \\[0.4cm]\displaystyle
=\sum_{t=0}^{\infty} \left( \frac{i \hbar}{2}\right)^t \frac{1}{t!}
\omega^{p_1 i_1} \cdots \omega^{p_t i_t} \frac{\partial^t
(b_{z,v_1 \ldots v_d,r_1 \ldots r_w} X^{v_1}
 \cdots X^{v_d}
)}{\partial X^{p_1} \ldots \partial X^{p_t}}   \cdot
\frac{\partial^t (a_{k,u_1 \ldots u_l,j_1 \ldots j_m} X^{u_1}
 \cdots X^{u_l}
)}{\partial X^{i_1} \ldots \partial X^{i_t}}
\end{array}\]
\be
=\sum_{t=0}^{\infty}
(-1)^t
 \left( \frac{i \hbar}{2}\right)^t
\frac{1}{t!} \omega^{i_1 p_1} \cdots \omega^{i_t p_t}
\frac{\partial^t (b_{z,v_1 \ldots v_d,r_1 \ldots r_w} X^{v_1}
 \cdots X^{v_d}
)}{\partial X^{p_1} \ldots \partial X^{p_t}}  \cdot
\label{g23}
\frac{\partial^t (a_{k,u_1 \ldots u_l,j_1 \ldots j_m} X^{u_1}
 \cdots X^{u_l}
)}{\partial X^{i_1} \ldots \partial X^{i_t}}. \ee 
Moreover, \be \label{g24} (dq^{r_1} \wedge \cdots \wedge dq^{r_w})
 \wedge
(dq^{j_1} \wedge \cdots \wedge dq^{j_m})= (-1)^{w \cdot m}
(dq^{j_1} \wedge \cdots \wedge dq^{j_m}) \wedge (dq^{r_1} \wedge
\cdots \wedge dq^{r_w}). \ee

Substituting (\ref{g23}) and (\ref{g24}) into the definition of
commutator $[a,b]$    (see \ref{16}) we obtain the relation
\be\label{d26}\begin{array}{l}
[a,b]=2
\sum_{t=0}^{\infty} \left( \frac{i \hbar}{2}\right)^{2t+1}
\frac{1}{(2t+1)!}
\omega^{i_1 p_1} \cdots \omega^{i_{2t+1} p_{2t+1}}
\frac{\partial^{2t+1} (a_{k,u_1 \ldots u_l,j_1 \ldots j_m} X^{u_1}
 \cdots X^{u_l}
)}{\partial X^{i_1} \ldots \partial X^{i_{2t+1}}} 
\\[0.3cm]
\quad \times  \frac{\partial^{2t+1} (b_{z,v_1 \ldots v_d,r_1
\ldots r_w} X^{v_1}
 \cdots X^{v_d}
)}{\partial X^{p_1} \ldots \partial X^{p_{2t+1}}} (dq^{j_1} \wedge
\cdots \wedge dq^{j_m}) \wedge (dq^{r_1} \wedge \cdots \wedge
dq^{r_w}) ,
\end{array}\ee 
so that only terms with odd number of derivatives in
$X^i $ appear in the commutator $[a,b].$ \rule{2mm}{2mm}

 The following corollaries are  conclusions  of the theorem
(\ref{tw21}):

\begin{co}
\label{co11} The commutator $[a,b] $ of two real forms $a \in
{\cal V} \otimes \Lambda^{m}$ and $
 b \in {\cal V} \otimes \Lambda^{w}$ is purely imaginary.
\end{co}

\begin{co}
\label{co12}
Assume that the $1-$form $\Gamma $ of the symplectic connection $ \nabla$ is real. Then 
the $1-$form $\tilde{\Gamma} $ of the abelian connection \[
\tilde{\Gamma}=\Gamma +\omega_{ij}X^i dq^j +r,
\]
 where $ r $ is defined by the formula (\ref{33})  with $ \delta^{-1}r=0, $ is also real.
\end{co}
\underline{Proof}

It is sufficient to show that  for   a real $1-$form   $ \Gamma $
the $1-$ form $ r $ is real    too.

From the relation (\ref{25}) and the theorem (\ref{tw21}), we can
see that the 2-form of the symplectic curvature $R $ is real for
the real  $ \Gamma. $ Using the linearity of the operator $
\delta^{-1}, $ the expression  (\ref{24}) of the exterior
covariant derivative $ \partial $ and the fact that $ r^2=
\frac{1}{2}[r,r], $ we can conclude that $r $ is real.
\rule{2mm}{2mm}

\begin{co}
\label{co13}
Assume that the $1-$form of  connection $\Gamma $ is real.
The 0-form  $ \sigma^{-1}(F) $ representing the real function $F$ is also real.
\end{co}

This property is a consequence of the formula (\ref{34}) of the
mapping  $ \sigma^{-1}(F), $ the corollary (\ref{co12}) and the
theorem (\ref{tw21}). \rule{2mm}{2mm}

\begin{tw}
\label{tw22} The real part of the Moyal product $ F_1*F_2$ of two
observables is symmetric and the imaginary part of $ F_1*F_2$ is
skew-symmetric under the interchange of the order of functions,
i.e. \be \label{1003}
 Re(F_1*F_2)=Re(F_2*F_1)   \qquad
 Im(F_1*F_2)=-Im(F_2*F_1). 
\ee 
\end{tw}

\noindent
\underline{Proof}

 From the definition (\ref{de00}) of an observable and the corollary
 (\ref{co13}),
$ 0 $--forms $ \sigma^{-1}(F_1) $ and $ \sigma^{-1}(F_2) $ are
real. In the product $F_1*F_2$ (from (\ref{35})), only elements $
\sigma^{-1}(F_1)  \circ \sigma^{-1}(F_2)$ not including $X^i$'s
are important.

Let us look on the multiplication of  two 0-forms
\be
\label{gr1}
a=  a_{k,u_1 \ldots u_l} X^{u_1}
 \cdots X^{u_l}, \qquad 
b=  b_{z,v_1 \ldots v_l} X^{v_1}
 \cdots X^{v_l},
\ee
which appear in
 $\sigma^{-1}(F_1)$ and   $\sigma^{-1}(F_2) $ respectively.
Now \be \label{01} \sigma(a \circ b)= \left( \frac{i
\hbar}{2}\right)^l \frac{1}{l!} \omega^{i_1 p_1} \cdots
\omega^{i_l p_l} \frac{\partial^l (a_{k,u_1 \ldots u_l} X^{u_1}
 \cdots X^{u_l}
)}{\partial X^{i_1} \ldots \partial X^{i_l}} \cdot
\frac{\partial^l (b_{z,v_1 \ldots v_l} X^{v_1}
 \cdots X^{v_l}
)}{\partial X^{p_1} \ldots \partial X^{p_l}}
\ee
and
\be
\label{02}
\sigma(b \circ a)=
\left( \frac{i \hbar}{2}\right)^l
\frac{1}{l!}
\omega^{i_1 p_1} \cdots \omega^{i_l p_l}
\frac{\partial^l (b_{z,u_1 \ldots u_l} X^{u_1}
 \cdots X^{u_l}
)}{\partial X^{i_1} \ldots \partial X^{i_l}} \cdot
\frac{\partial^l (a_{k,v_1 \ldots v_l} X^{v_1}
 \cdots X^{v_l}
)}{\partial X^{p_1} \ldots \partial X^{p_l}}. \ee Changing the
indices $u_s \leftrightarrow v_s $ in the formula (\ref{02}) we
obtain \be \label{03} \sigma(b \circ a)= \left( \frac{i
\hbar}{2}\right)^l \frac{1}{l!} \omega^{i_1 p_1} \cdots
\omega^{i_l p_l} \frac{\partial^l (b_{z,v_1 \ldots v_l} X^{v_1}
 \cdots X^{v_l}
)}{\partial X^{i_1} \ldots \partial X^{i_l}} \cdot
\frac{\partial^l (a_{k,u_1 \ldots u_l} X^{u_1}
 \cdots X^{u_l}
)}{\partial X^{p_1} \ldots \partial X^{p_l}}. \ee But $\omega^{i_1
p_1}=- \omega^{p_1 i_1},$ so that from (\ref{01}) and (\ref{03})
we can see that \be \label{04} \sigma(a \circ b)= (-1)^l \sigma(b
\circ a). \ee If $ l $ is even, the element $\sigma(a \circ b) $
is real, for $ l $ odd we have that $\sigma(a \circ b) $ is
imaginary. This fact along (\ref{04}) gives us the proof of the
theorem \ref{tw22}. \rule{2mm}{2mm}

\sect{Quantization of a $2$--D phase space with constant curvature tensor}

 Using the information contained in previous sections, we
shall accomplish the main task of this contribution, the
quantization of momentum and position on the $ 2$--D
phase space $ T^*\mathbb{R} $ with a constant curvature tensor.


\subsect{Symplectic connection $\nabla $ and abelian connection 
$\tilde{\nabla}$ on  $ T^*\mathbb{R} $}
\label{abe}

Let $\mathbb{R}$ be our configuration space.
 We cover the configuration space by the atlas $ \{(\mathbb{R},
\tilde{\varrho}) \} $ consisting of  one chart $ (\mathbb{R},
\tilde{\varrho}). $ The Riemannian connection characterized by the
coefficient $ \Gamma^1_{11} $ vanishes. In the atlas $
\{(\mathbb{R}, \tilde{\varrho}) \}, $
 information about dynamics of the system is given by the
 vector from the cotangent space $ T^*_{\tt p}(\mathbb{R}) $
\be
\label{201}
[{\tt p},(\mathbb{R}, \tilde{\varrho}), p],
\ee
where $ p $ is the momentum coordinate at the point $ {\tt p} \in \mathbb{R} $ in the natural
basis  defined by the chart  $ (\mathbb{R}, \tilde{\varrho}). $
We do not put any constraints on the system, hence its
phase space is the cotangent bundle
\be
\label{202}
 T^*\mathbb{R} = \bigcup_{{\tt p} \in \mathbb{R}} T^*_{\tt p}(\mathbb{R}).
\ee
 This $ 2$--D space is equipped with some symplectic structure $ \omega . $  The atlas
$
\{(\mathbb{R}, \tilde{\varrho}) \} $ on the configuration space determines an  atlas $
\{(\mathbb{R}^2, \varrho) \} $ on the space
$ T^*\mathbb{R}. $ The diffeomorphism $ \varrho $ is defined by
\be \label{203} \varrho({\tt p}):=(q,p) \qquad \forall {\tt p} \in
T^*\mathbb{R}. \ee In the chart $ (\mathbb{R}^2, \varrho) $ the
symplectic form is \be \label{2031}
 \omega =dp \wedge dq
\ee
and the  Poisson tensor
\be
\label{Poisson}
\omega^{ij}=
\left(\begin{array}{cc}
0 &1 \\
- 1&0
\end{array} \right).
\ee The atlas $ \{(\mathbb{R}^2, \varrho) \} $ is an example of
the proper Darboux atlas (see \cite{3}). Coordinates $ (q,p) $ are
called {\it proper Darboux coordinates} or {\it induced
coordinates}. In all the charts in
 a proper Darboux atlas on a symplectic manifold
$ T^*{\cal M} \;\;\; (\dim \; T^*{\cal M}=2n), $ the coordinates $
q^i, \;\;\; 1 \leq i \leq n, $ determine points on the basic
manifold $ {\cal M} $ and $ q^{i+n}=p_i \;\;\; 1 \leq i \leq n, $
denote momenta in  natural coordinates. The transition functions
are the point transformations \be \label{20305} Q^j=Q^j(q^i)
\qquad, \qquad P_j= \frac{\partial q^i}{\partial Q^j}p_i. \ee The
relations (\ref{20305}) define {\it proper Darboux
transformations.} Proper Darboux charts preserve the  difference,
obvious from physical point of view, between spatial coordinates
and momenta. Topologically, the space $ T^*\mathbb{R} $ is
homeomorphic to $ \mathbb{R}^2. $ Although $ T^*\mathbb{R} $ is
also equipped with some additional structure: the symplectic
form. Moreover,  we do not define any metric structure on the
phase space.  With the form $ \omega $ defined as (\ref{2031}), we
connect the geometric structure on $ T^*\mathbb{R} $
characterized by some connection, its torsion and curvature, which
will be different from the Riemannian structure of $ \mathbb{R}^2.
$ This is the reason why there is no sense to talk about the shape
of $ T^*\mathbb{R} $ and to analyze its properties as features of
some manifold embedded in the metric space $ \mathbb{R}^3. $   As
we mentioned in the previous section, the symplectic connection is
not defined uniquely.  In this paper we use the method for the
construction of the symplectic connection presented in \cite{3}.
Another choice  has been proposed in \cite{prz}.   The index `1'
is assigned to the spatial coordinate $ q $ and `2' to the
momentum  $ p. $  The construction is performed in the atlas $
\{(\mathbb{R}^2, \varrho) \}. $ The coefficient $ \Gamma_{211} $
vanishes because this element is determined by the Riemannian
connection on the configuration space and we have chosen the chart
$ (\mathbb{R}, \tilde{\varrho}) $ in such a way that $
\Gamma^1_{11}=0. $ The coefficients $ \Gamma_{122} $ and $
\Gamma_{222} $ transform like tensors under proper Darboux
transformations and there are not any extra restrictions on them,
hence they are chosen to be zero. There is no reasonable physical
restrictions on the form of the coefficient $ \Gamma_{111}. $  The
case $ \Gamma_{111}=0 $ is  trivial. We are going to analyze the
situation when $ \Gamma_{111}=p $,  i.e., this coefficient is a
linear function of $ p. $ The manifold $ T^*\mathbb{R} $ is
covered by the one chart $ (\mathbb{R}^2, \varrho) $, so we need
not use a partition of unity to define  $ \Gamma_{111} $ globally.

The 1-form of the connection can be written \be \label{204}
\Gamma= \frac{1}{2} p X^1 X^1 dq. \ee

The 2-form of the curvature is given by \be \label{205}
R=-\frac{1}{2}X^1 X^1 dq \wedge dp ,
\ee 
so that, the phase space $
T^* \mathbb{R} $ has constant symplectic curvature. The
nonvanishing elements of the symplectic curvature tensor are \be
\label{205a} R_{1112}= -R_{1121}= - \frac{1}{4}. \ee The Ricci
tensor on a symplectic manifold is defined as \be \label{205b}
K_{ij}:=\omega^{kl}R_{likj} \ee and in our case the only one
nontrivial component of $K_{ij}$ is the term \be \label{205c}
K_{11}= \frac{1}{4}. \ee The scalar of curvature is given by the
contraction \be \label{205d} K := \omega^{ji}K_{ij}, \ee which is
equal zero. This result is true for any symplectic manifold (see
\cite{Va1}). Having the 1-form of the connection defined by the
expression (\ref{204}), we are able to build the abelian
connection. At the beginning, we compute the series $ r $ defined
by (\ref{33}).  The first term in the recurrent expression
(\ref{33}) is \be \label{206} \delta^{-1}R = \frac{1}{8} X^1 X^1
X^2 dq - \frac{1}{8} X^1 X^1 X^1 dp. \ee Using the theorem
(\ref{tw21}) and the formula (\ref{33}) we  can see that in the
series  $ r $ appear  two kind of terms only:
\[
(-1)^{m+1} a_{2m+1} (X^1)^{2m}X^2 dq, 
\qquad
(-1)^m a_{2m+1} (X^1)^{2m+1} dp, \qquad  m \in {\mathbb N}-\{0\}.
\]
 The element $ a_3, $
being the starting point for $a_{2m+1} \in \mathbb{R}$, is
determined by $ \delta^{-1}R $ and from (\ref{206}) equals to \be
\label{207a}
 a_3 = \frac{1}{8}.
\ee It is easy to check that the coefficients $a_{2m+1}$ for $ m >
1 $ are given by the recurrent formula \be \label{207}
a_{2m+1}=\frac{1}{2m+2} \sum_{g=1}^{m-1}(2g+1) \cdot a_{2g+1}
\cdot a_{2m-2g+1}. \ee The explicit form of $ a_{2m+1} $ is
related to the so called  Catalan numbers and it is given by the
equation  
\be \label{207q} a_{2m+1} =
\frac{2}{m} \left(
\begin{array}{c}
2m-2 \\
m-1
\end{array}
\right) \frac{1}{16^m} \qquad {\rm for}\qquad m \geq 1. 
\ee
Remembering  that the abelian connection  $ \tilde{\Gamma} $ is a
sum of three expressions (see (\ref{32})): $ \Gamma, r  $ and $
-X^1dp + X^2 dq $ we obtain  the following relation 
\be
\label{208} 
\begin{array}{l}
 \tilde{\Gamma}= -X^1dp + X^2 dq + \frac{1}{2} p X^1
X^1 dq  -\frac{1}{8} X^1 X^1 X^1 dp
   +\frac{1}{8} X^1 X^1 X^2 dq \\[0.3cm]
\qquad 
+\frac{1}{128}  X^1 X^1 X^1 X^1 X^1 dp
-\frac{1}{128}  X^1 X^1 X^1 X^1 X^2 dq  \\[0.3cm]
\qquad - \frac{1}{1024} X^1 X^1 X^1 X^1 X^1 X^1 X^1 dp + \frac{1}{1024}
X^1 X^1 X^1 X^1 X^1 X^1 X^2 dq - \cdots 
\end{array} \ee
 The abelian connection is an infinite series. In none of the terms
belonging to $ \tilde{\Gamma} $ the Planck constant $ \hbar $
appears.

\subsect{The eigenvalue equation for momentum $ p $ }

In the previous section, we found the
abelian connection on the symplectic space $ T^*\mathbb{R} $ with
constant curvature tensor. The expression (\ref{208}) enables us
to write eigenvalue equations for  observables.
 Let us start with constructing the explicit form of the eigenvalue equation 
for momentum $ p. $
The general form of this equation is
\be
\label{31u}
p * W_{\bf p }(q,p)= {\bf p } \cdot W_{\bf p }(q,p),
\ee
where $ {\bf p } $ denotes the eigenvalue of $p$ and $ W_{\bf p }(q,p)$ is 
the Wigner function for the eigenvalue $ {\bf p }.
$
 Using the iterative relation (\ref{34}) we find the general formula for
$ \sigma^{-1}(p). $  After a simple but rather long computation,
we arrive to the expression 
\be
\label{310} 
\begin{array}{lll}
 \sigma^{-1}(p) &=&  p +
X^2 + \frac{1}{2} p X^1 X^1 +\frac{1}{8} X^1 X^1 X^2 \\[0.3cm]
& & \quad   -\frac{1}{128}  X^1 X^1 X^1 X^1 X^2 +
\frac{1}{1024} X^1 X^1 X^1 X^1 X^1 X^1 X^2  - \cdots
\end{array} \ee
Every
coefficient $ b_{2m+1}$ appearing in the term $
b_{2m+1}\underbrace{X^1 \cdots X^1 }_{2m }X^2$ for $ m > 0 $ can
be expressed in the form \be \label{309}
b_{2m+1}=(-1)^{m+1}a_{2m+1}, \ee where the numbers $ a_{2m+1} $
are defined by  (\ref{207}) with the initial condition
(\ref{207a}).

Much more complicated is the problem of finding the general formula of 
the series $\sigma^{-1}(W_{\bf p })$
representing the Wigner function $W_{\bf p }(q,p).$

Let us start from the computation of  terms including $ i
\hbar^{2s+1} $ in the eigenvalue equation (\ref{31u}). Using
(\ref{310}), we deduce that
\be
\label{311} 
\begin{array}{l}
(i \hbar)^{2s+1} A_{2s+1}(q,p)= \sigma \left(
b_{2s+1}(X_1)^{2s}X_2 \circ w_{0,1,2s} X^1(X^2)^{2s} \right. \\[0.3cm]
\qquad \left. +\;  b_{2s-1}(X^1)^{2s-2}X^2 \circ  \hbar^2
w_{2,1,2s-2} X^1(X^2)^{2s-2} + \cdots  + X^2 \circ
\hbar^{2s}w_{2s,1,0}X^1 \right) .
\end{array} \ee
 The functions
$A_{s}(q,p)$ are the coefficients of  the odd powers of $\hbar
$ in the expansion   of $p * W_{\bf p}(q,p).$
The coefficients
 $ b_{s} $  are
defined in (\ref{309}).
The numbers
 $ w_{a,b,c}$ are the  coefficients of $\hbar^a (X^1)^b (X^2)^c$ in the span of 
$ \sigma^{-1}(W_{\bf p }(p,q)). $

For $s \geq 1, $
\be
\label{312} 
\begin{array}{lll} 
\hbar^{2s}w_{2s,1,0}X^1 &=&  \left(\displaystyle
\frac{\hbar}{2}
\right)^{2s}(-1)^{s+1}\displaystyle\frac{(2s)!}{2s+1}b_{2s+1}w_{0,1,2s}X^1 + \cdots
\\[0.3cm]
& & \quad + \left(\displaystyle\frac{\hbar}{2}
 \right)^{2s}(-1)^{1+1}\displaystyle\frac{(2 \cdot 1)!}{2\cdot 1
+1}b_{3}w_{2s-2,1,2}X^1.
\end{array} \ee
 Carrying (\ref{312})  into (\ref{311}) we obtain
\be \label{313} A_{2s+1}(q,p)=0 \;\; {\rm for} \;\; s \geq 1. \ee
It means that the only  complex term appearing in the product $p
* W_{\bf p }(q,p)$ is the element
\be \label{314} i \hbar
A_{1}(q,p) = i \hbar \frac{\partial W_{\bf p }(q,p)}{\partial q}.
\ee
The proof of (\ref{313}) and (\ref{314})  can be found in  Appendix
 A.
The Wigner function must be real, hence    the element $ A_{1}(q,p) $
vanishes.

We conclude that the function $W_{\bf p }(q,p)$ depends only on $p. $
Instead of looking for the general form of the expression $ \sigma^{-1}(W_{\bf p }(q,p)),$
we are interested in the formula for $ \sigma^{-1}(W_{\bf p }(p)).$ In 
$ \sigma^{-1}(W_{\bf p }(p))$
 only terms with even power of $ \hbar $ appear and they can be written in the recurrent
form \be \label{314aa} w_{2k,0,c}= \frac{1}{c} \frac{d
w_{2k,0,c-1}}{d p} + \sum_{t=1}^{k} \frac{1}{4^t}
\frac{(2t+c)!}{c!}a_{2t+1}w_{2k-2t,0,2t+c}, \qquad k,c \geq 0. \ee
The coefficients $ a_{2t+1} $ are defined by   (\ref{207a}) and
(\ref{207}). For $ k=0, $ we put the sum in (\ref{314aa}) equal $
0. $ Moreover, it is possible to prove that the projection $
\sigma $ keeps from the product \\ $ \sigma^{-1}(p) \circ
\sigma^{-1}(W_{\bf p }(p)) $ only terms of the form \be
\label{314a}
 \sigma \left( \frac{1}{2} p X^1 X^1 \circ w_{2k,0,2} \hbar^{2k} X^2 X^2 \right)
= -\frac{\hbar^{2k+2}}{4} p \: w_{2k,0,2}.
\ee

Taking into account (\ref{314}) and (\ref{314a}), we arrive to the conclusion
that the eigenvalue equation for $W_{\bf p }(p)$ is the infinite
differential equation
\be
\label{315}
p \cdot \left(
W_{\bf p }(p) - \frac{1}{8}\hbar^2 \frac{d^2 W_{\bf p }(p)}{dp^2}
- \frac{1}{128}\hbar^4 \frac{d^4 W_{\bf p }(p)}{dp^4}  - \cdots
\right)= {\bf p } \cdot
 W_{\bf p }(p).
\ee
 Notice that, although the product $ p * W_{\bf p}(p) $ is abelian, 
the expression on the left side of the equation (\ref{315}) is not simply  $ p
\cdot  W_{\bf p}(p). $

The eigenvalue equation for momentum $ p $ is a differential equation of  infinite degree.
There is no general method of solving such equations. Therefore,
 we decided to look for the solution of the  eigenvalue equation of $p^2.$ As
$
p \cdot p= p * p \;
$
 (which is not true for  $p \cdot p \cdot p \neq p*p*p $),
 the Wigner function fulfilling the equation (\ref{31}) satisfies also the
relation
\be
\label{316}
p^2*W_{\bf p }(p)= {\bf p^2 } W_{\bf p }(p).
\ee
After a long but elementary computation (see Appendix B), we can see that 
the above relation is the modified Bessel equation
\be
\label{317}
\frac{1}{4}\hbar^2 p^2 \frac{d^2 W_{\bf p }(p)}{dp^2}
+\frac{1}{4}\hbar^2 p \frac{d W_{\bf p }(p)}{dp}
+({\bf p^2 }-p^2)W_{\bf p }(p)=0.
\ee
The general solution  of (\ref{317}) is a linear combination  of the following form
\be
\label{318}
W_{\bf p }(p)=A \cdot I_{\frac{2i{\bf p }}{\hbar}}(\frac{2p}{\hbar})
+ B \cdot K_{\frac{2i{\bf p }}{\hbar}}(\frac{2p}{\hbar})
\ee
where $I_{\frac{2i{\bf p }}{\hbar}}(\frac{2p}{\hbar})
$ is the modified Bessel function with  complex parameter $\frac{2i{\bf p }}{\hbar}$
and $ K_{\frac{2i{\bf p }}{\hbar}}(\frac{2p}{\hbar})$ is the modified Bessel 
function of the second kind with
 parameter $\frac{2i{\bf p }}{\hbar}.$ This solution is defined for arguments 
$ \frac{2p}{\hbar}>0 $ (see \cite{Sch}).

Notice that the deformation parameter $ \hbar $ appears   in the denominator of the argument. 
We must be very careful
because this fact may cause the
 nonexistence of the series  $ \sigma^{-1}(W_{\bf
p}(p)). $

Until this moment, it was not necessary to use any explicit form of
the functional action  $ \langle W(q,p),F(q,p)\rangle $ where $
F(q,p) $ is some function on $ T^*\mathbb{R}. $ In the general
case \be \label{318a} \langle W(q,p),F(q,p)\rangle :=
\int_{T^*\mathbb{R}} W(q,p)
* F(q,p) t(q,p)dqdp. \ee From theorem \ref{wazne} we can see that
in our example the trace density $t(q,p)$ is constant, so we can
write \be t(q,p)=1. \ee The straightforward consequence of this
fact is that the $*$--product used by us is closed  (\cite{jap},
\cite{con}).

Let us come back to the equation (\ref{318}). The function $ A
\cdot I_{\frac{2i{\bf p }}{\hbar}}(\frac{2p}{\hbar}) $ is
complex. Its real part grows up to infinity for $ x \rightarrow
\infty $ and hence it is not normalizable. The imaginary part of $
I_{\frac{2i{\bf p }}{\hbar}}(\frac{2p}{\hbar})$  is proportional
to $ K_{\frac{2i{\bf p }}{\hbar}}(\frac{2p}{\hbar}).$ Because of
these reasons we affirm that the only  physically acceptable
solution of (\ref{317}) is \be \label{319} W_{\bf p }(p)= B \cdot
K_{\frac{2i{\bf p }}{\hbar}}(\frac{2p}{\hbar}). \ee However,
solutions defined on the whole $ \mathbb{R} $ are required. It is
impossible to define the solution of the equation (\ref{317}) on
the whole axis. The problem is that the modified Bessel function
of the second kind is not defined for the value of the argument
$p=0.$ Moreover, $\lim_{p\rightarrow 0^+} K_{\frac{2i{\bf p
}}{\hbar}}(\frac{2p}{\hbar}) $ does not exist.

Let us assume that the Wigner function $ W_{\bf p }(p) $ is some
generalized function over the Schwartz space $ {\cal S}(p) $ of
test smooth functions tending to $0$ for $ p \rightarrow \pm
\infty$ faster than the inverse of any polynomial.

We define the Wigner function $W_{\bf p }(p)$ as:
\begin{enumerate}
\item
 for ${\bf p}<0$
we put
\be
\label{aa}
 W_{\bf p} (p)=
\left\{
\begin{array}{cc}
 \frac{4 }{\pi \hbar} \cosh{\frac{{\bf p} \pi }{\hbar}}
K_{\frac{2i{\bf p }}{\hbar}}(\frac{-2p}{\hbar}) &
\mbox{for}
\;\;\; p<0 \\[0.3cm]
0  & \mbox{for} \;\;\; p \geq 0 ;
\end{array}
\right.
\ee

\item
For ${\bf p}>0,$ we have \be \label{bb}
 W_{\bf p} (p)=
\left\{
\begin{array}{cc}
0  & \mbox{for} \;\;\; p \leq 0\\[0.3cm]
\frac{4 }{\pi \hbar} \cosh{\frac{{\bf p} \pi }{\hbar}}
K_{\frac{2i{\bf p }}{\hbar}}(\frac{2p}{\hbar}) & \mbox{for}
\;\;\; p>0
\end{array}
\right.
\ee

\item
and finally for ${\bf p}=0$: \be \label{cc}
 W_{\bf 0} (p)=
\left\{
\begin{array}{cc}
\frac{2 }{\pi \hbar}
K_{0}(\frac{-2p}{\hbar})
& \mbox{for} \;\;\; p<0\\[0.3cm]
\frac{2 }{\pi \hbar}
K_{0}(\frac{2p}{\hbar}) & \mbox{for}
\;\;\; p>0.
\end{array}
\right.
\ee

\end{enumerate}

The coefficients $ \frac{4 }{\pi \hbar} \cosh{\frac{{\bf p} \pi
}{\hbar}} $ or $\frac{2 }{\pi \hbar} \cosh{\frac{{\bf p} \pi
}{\hbar}} $ appear as the result of normalization, i.e. we require
that \be \label{320a} \int_{-\infty}^{\infty}1 \cdot W_{\bf p}(p)
dp=1. \ee

The reason for not integrating over $dq$ is that our Wigner
function does not depend on $q. $  We can see that the Wigner
eigenfunction of momentum $ p $ is a locally summable function.

 As we mentioned before for ${\bf p} \neq 0 $ there is no
$\lim_{p\rightarrow 0^+} K_{\frac{2i{\bf p
}}{\hbar}}(\frac{2p}{\hbar}). $ However, the functions
\\$ \frac{4 }{\pi \hbar} \cosh{\frac{{\bf p} \pi }{\hbar}}K_{\frac{2i{\bf
p }}{\hbar}}(\frac{-2p}{\hbar}) $ and
$ \frac{4 }{\pi \hbar} \cosh{\frac{{\bf p} \pi }{\hbar}}
K_{\frac{2i{\bf p }}{\hbar}}(\frac{2p}{\hbar}) $ are bounded. 
The consequence of this fact is that (\ref{aa}) and (\ref{bb}) are solutions of (\ref{317}).

Indeed, let ${\bf p}>0.$
For every test function $\varphi(p) \in {\cal S}(p)$ we have
\[
\langle\langle p \frac{dW_{\bf p}(p)}{dp},\varphi(p)
\rangle\rangle= \int_{0}^{\infty} p \frac{4 }{\pi \hbar}
\cosh\left({\frac{{\bf p} \pi }{\hbar}}\right) \frac{d}{dp} \left(
K_{\frac{2i{\bf p }}{\hbar}}(\frac{2p}{\hbar}) \right)  \varphi(p)
dp.
\]
Integrating the above expression by parts we obtain
\[\begin{array}{lll}
  \langle\langle p \displaystyle\frac{dW_{\bf p}(p)}{dp},\varphi(p) \rangle\rangle &=&
 p \frac{4 }{\pi \hbar}\displaystyle
\cosh\left({\frac{{\bf p} \pi }{\hbar}}\right) K_{\frac{2i{\bf p
}}{\hbar}}(\frac{2p}{\hbar}) \varphi(p) |_{0}^{\infty}
\\[0.3cm]
& & \quad- \displaystyle\int_{0}^{\infty}\frac{4 }{\pi \hbar} \cosh\left({\frac{{\bf p} \pi
}{\hbar}}\right) K_{\frac{2i{\bf p }}{\hbar}}(\frac{2p}{\hbar})\cdot
\left( \varphi + p \frac{d \varphi(p)}{dp}\right)dp.
\end{array}\]
The expression $ p \frac{4 }{\pi \hbar} \cosh\left({\frac{{\bf p} \pi }
{\hbar}}\right) K_{\frac{2i{\bf p }}{\hbar}}(\frac{2p}{\hbar})
 \varphi(p) |_{0}^{\infty}$ vanishes. Thus,
 we can see that as  generalized function 
\be \label{ma1} p
\frac{dW_{\bf p}(p)}{dp}= \left\{
\begin{array}{cc}
0  & \mbox{for} \;\;\; p \leq 0\\[0.3cm]
p \frac{4 }{\pi \hbar} \cosh{\frac{{\bf p} \pi }{\hbar}} \frac{d}{dp} \left(
K_{\frac{2i{\bf p }}{\hbar}}(\frac{2p}{\hbar}) \right) & \mbox{for}
\;\;\; p>0.
\end{array}
\right. \ee Analogously, we can show  that \be \label{ma2} p^2
\frac{d^2 W_{\bf p}(p)}{dp^2} = \left\{
\begin{array}{cc}
0  & \mbox{for} \;\;\; p \leq 0\\[0.3cm]
p^2 \frac{4 }{\pi \hbar} \cosh{\frac{{\bf p} \pi }{\hbar}} \frac{d^2}{dp^2}\left(
K_{\frac{2i{\bf p }}{\hbar}}(\frac{2p}{\hbar}) \right) & \mbox{for}
\;\;\; p>0.
\end{array}
\right.
\ee

Putting (\ref{ma1}) and (\ref{ma2}) into (\ref{317}), we arrive to
the conclusion that (\ref{bb}) really fulfills (\ref{317}).
 For ${\bf p} <0$
 the proof is analogous.

If $ {\bf p} =0 $ we have the limit
\[
\lim_{p\rightarrow 0^+} K_{0}(\frac{2p}{\hbar})= \infty
\]
but
\[
\lim_{p\rightarrow 0^+}p K_{0}(\frac{2p}{\hbar})= 0
\]
and
\[
\lim_{p\rightarrow 0^+}p^2 \frac{d}{dp} K_{0}(\frac{2p}{\hbar})= 0.
\]

Thus following the way presented in the above proof, we see that
also (\ref{cc}) is the solution of (\ref{317}) for $ {\bf p} =0. $

To be precise, at this moment we only know that the generalized
functions (\ref{aa})--(\ref{cc}) are solutions of (\ref{317}). To
give a physical interpretation to (\ref{aa})--(\ref{cc}), we must
prove that series
 $ \sigma^{-1}\left( W_{\bf p}(p) \right) $ exist for every 
${\bf p}$ and that they fulfill not only (\ref{317}) but also
(\ref{315}).  The general expression for $ \sigma^{-1}\left(
W_{\bf p}(p) \right) $ for every of the solutions 
(\ref{aa})--(\ref{cc}) is very complicated so we concentrate only on those
terms appearing in the product $  \sigma^{-1}(p) \circ
\sigma^{-1}\left( W_{\bf p}(p) \right). $ As we have shown, the
expression for  $ w_{2k,0,2} $ is defined by (\ref{314aa}).
Introducing a new variable $ z = \frac{2p}{\hbar},$ we can see that
the coefficient of $ X^2 X^2 $ equals to \be \label{?1}
  \sum_{k=1}^{\infty}4^k a_{2k+1}\frac{d^{2k} W_{\bf p}(z)}{dz^{2k}}.
\ee We  first consider the case ${\bf p} = 0.$ From (\ref{207q})
we can see that it is necessary to analyze the expression \be
\label{??1}
 T W_{\bf 0}(z) := \sum_{m=1}^{\infty}\frac{2}{m} \left(
\begin{array}{c}
2m-2 \\
m-1
\end{array}
\right) \frac{1}{4^m}\frac{d^{2m} W_{\bf 0}(z)}{dz^{2m}}. \ee We
know that $W_{\bf 0}(z)$ is a tempered distribution and we would
like to prove that so is $ T W_{\bf 0}(z). $ To do that, we use
the Fourier transform. To avoid any misunderstandings we remind
the basic definitions and results connecting with the Fourier
transform (compare \cite{sw}).

For every $ \varphi(p) \in {\cal S}(p) $ \be \label{??2} {\cal
F}[\varphi](\lambda) := \int_{- \infty}^{+ \infty} e^{- 2 \pi i
\lambda p} \varphi(p) dp. \ee The Fourier transform $ {\cal
F}[U](\lambda) $ of a tempered distribution is a tempered
distribution fulfilling the equation \be \label{??3}  \langle
{\cal F}[U],\varphi \rangle =  \langle U,{\cal F}[\varphi] \rangle,
\ee for every test function $\varphi(\lambda) \in {\cal
S}(\lambda).$ The consequences of (\ref{??2}) and (\ref{??3}) are
the following relations 
\be \label{??4} {\cal F}[\frac{d^m
U}{dp^m}]= (2 \pi i \lambda)^m {\cal F}[U], \ee \be \label{??5}
{\cal F}[\delta ] = 1, \ee \be \label{??6} {\cal F}[ \frac{d^m
\delta}{dp^m} ] = (2 \pi i \lambda)^m. \ee

Let us compute now the Fourier transform of (\ref{??1}). Using
(\ref{??4}) we obtain \be \label{??7} {\cal F}[T W_{\bf 0}(z)]=
\left( 1- \sum_{m=1}^{\infty}\frac{2}{m} \left(
\begin{array}{c}
2m-2 \\
m-1
\end{array}
\right) \frac{1}{4^m} (2 \pi i \lambda)^{2m} \right) {\cal F} [ W_{\bf 0}(z)].
\ee
The infinite power series 
\[ 1-
\sum_{m=1}^{\infty}\frac{2}{m} \left(
\begin{array}{c}
2m-2 \\
m-1
\end{array}
\right) \frac{1}{4^m} (2 \pi i \lambda)^{2m}
\] 
is the MacLaurin
series of the function 
\be \label{??8} \sqrt{1+(2 \pi \lambda)^2}.
\ee 
Hence, formally \be \label{??9} {\cal F}[T W_{\bf 0}(z)] =
\sqrt{1+(2 \pi \lambda)^2} {\cal F} [ W_{\bf 0}(z)]. \ee
Also,
 we know  that \cite{Gra} \be \label{??10} {\cal F} [
W_{\bf 0}(z)] =  \frac{1}{\sqrt{1+(2 \pi \lambda)^2}} \ee which finally gives that
 \be \label{??11} {\cal F} [T W_{\bf 0}(z)]=1. \ee With
(\ref{??5}) we conclude that \be T W_{\bf 0}(z) = \delta(z). \ee We have
proved that $T W_{\bf 0}(z) $ is really a tempered distribution.
It is obvious that $ \frac{2}{\hbar} \delta(\frac{2p}{\hbar}) $ is
the solution of (\ref{315}) for ${\bf p}=0$ and we remember that
the equation (\ref{315}) is a straightforward consequence of the
fact, that the coefficient at $X^2 X^2$ is equal to (\ref{??1}).

Doing similar computations, it is possible to prove that  $ T W_{\bf p}(z) $ 
is a tempered distribution also for $ {\bf p} \neq 0$ and that
the equation  (\ref{315}) is fulfilled.
Moreover, in those cases the relation  (\ref{315}) may be interpreted as 
a regularization of the distribution $ T W_{\bf p}(z). $

We were able to use the Fourier transform because the cotangent 
space $T_Q^*\mathbb{R}$ in every point $Q $
belonging to the configuration space is a metric space.

The constant  and negative curvature  on the phase space $
T^*\mathbb{R} $ divides this space in two separable subspaces:
 with positive values of the momentum and with negative ones respectively. 
The Wigner eigenvalue functions $ W_{\bf p} $
 with positive ${\bf p}$ vanish for $p<0$ and with
negative
$ {\bf p}$ $W_{\bf p}(p)$ is zero for $p>0.$ Only the solution for the eigenvalue ${\bf p}=0$ 
is defined on the whole axis $\mathbb{R}.
$
 The Wigner function $ W_{\bf p}(p) $ depends only on the variable $p$, but it is not
a distribution function in statistical sense  \cite{cb}. For some positively defined 
observables $F(p)$
\[
\langle W_{\bf p}(p),F(p) \rangle <0.
\]

The Wigner eigenfunctions (\ref{aa})--(\ref{cc}) fulfill
orthogonality relations \be \label{uhu} \langle\langle W_{\bf
p_1}(p)*W_{\bf p}(p),\varphi({\bf p},p) \rangle\rangle = \frac{1}{2
\pi \hbar}\int_{-\infty}^{+\infty} \varphi({\bf p_1},p)dp \ee for
every $\varphi({\bf p},p) \in {\cal S}^2({\bf p},p).$

The behavior of the Wigner eigenfunctions for ${\bf p}\neq 0$ on both
sides  of the point $0$ seems to be completely different, but in
fact due to fast oscillations of the Bessel functions in a
neighborhood of $0 $ their contribution to integrals
\[
\int_{-\infty}^{+\infty}W_{\bf p}(p)\varphi(p)dp
\]
is negligible.

The example analyzed before is a particular choice of $G$ in
 the $ 2 $--form of curvature
\be
\label{ex1}
R=-G^2 X^1X^1 dq \wedge dp,
\ee
where $ G$ is some positive constant.

In such a case the spectrum of the observable $ p $ is the whole axis $ \mathbb{R} $ 
and the Wigner functions are generalized functions:
\begin{enumerate}
\addtocounter{enumi}{3}
\item
 for ${\bf p}<0$
\be
\label{e1aa}
 W_{\bf p} (p)=
\left\{
\begin{array}{cc}
 \frac{4 }{G \pi  \hbar} \cosh{\frac{{\bf p} \pi }{ G \hbar}}K_{\frac{2i{\bf p }}
{ G\hbar}}(\frac{-2p}{G\hbar}) & \mbox{for}
\;\;\; p<0 \\[0.3cm]
0  & \mbox{for} \;\;\; p \geq 0 ;
\end{array}
\right.
\ee
\item
for ${\bf p}>0$
we have
\be
\label{e1bb}
 W_{\bf p} (p)=
\left\{
\begin{array}{cc}
0  & \mbox{for} \;\;\; p \leq 0\\[0.3cm]
\frac{4 }{G \pi  \hbar} \cosh{\frac{{\bf p} \pi }{ G \hbar}}
K_{\frac{2i{\bf p }}{G \hbar}}(\frac{2p}{ G \hbar}) & \mbox{for}
\;\;\; p>0 ;
\end{array}
\right.
\ee
\item

and  for ${\bf p}=0$ we put
\be
\label{e1cc}
 W_{\bf 0} (p)=
\left\{
\begin{array}{cc}
\frac{2 }{G \pi  \hbar}
K_{0}(\frac{-2p}{G \hbar})
& \mbox{for} \;\;\; p<0\\[0.3cm]
\frac{2 }{G \pi  \hbar}
K_{0}(\frac{2p}{G \hbar}) & \mbox{for}
\;\;\; p>0.
\end{array}
\right.
\ee
\end{enumerate}

For $ G \rightarrow 0^+ $ all of these solutions tend to $
\delta(p-{\bf p})$  (for details see the Appendix C). This result
is natural because for $ G=0 $ the phase space $ T^*\mathbb{R} $ is simply
$\mathbb{R}^2. $

In a similar way we deal with  the space $T^*\mathbb{R} $ equipped with the curvature form
\be
\label{ex2}
R = U^2 X^1X^1 dq \wedge dp,
\ee
where $U, $ as before, is some positive constant.

The eigenvalue equation for $p^2$ is
\be
p^2 W_{\bf p}(p) +
    \frac{1}{4}U^2\hbar^2 p \frac{d W_{\bf p}(p)}{dp}
  +
    \frac{1}{4}U^2 \hbar^2 p^2
\frac{d^2 W_{\bf p}(p)}{dp^2}= {\bf p}^2 W_{\bf p}(p).
\ee
The general solution of it is
\be
W_{\bf p}(p) = A \cdot J_{\frac{2{\bf p}}{U\hbar}}\left( \frac{2p}{U\hbar}\right)
 + B \cdot  Y_{\frac{2{\bf p}}{U\hbar}}\left( \frac{2p}{U\hbar}\right)
\ee
where $A, B$ are some real numbers and  $J_{\frac{2{\bf p}}{U\hbar}}
\left( \frac{2p}{U\hbar}\right)$
 and
$Y_{\frac{2{\bf p}}{U\hbar}}\left( \frac{2p}{U\hbar}\right)$ are the Bessel
 functions of the first and second kind, respectively.

It seems that every real number is the eigenvalue of momentum $ p $ 
and the Wigner eigenfunctions are generalized
functions
\begin{enumerate}
\setcounter{enumi}{6}
\item
 for ${\bf p}<0$
\be
\label{e2aa}
 W_{\bf p} (p)=
\left\{
\begin{array}{cc}
 \frac{2 }{ U \hbar}  J_{\frac{2|{\bf p }|}{ U\hbar}}(\frac{-2p}{U\hbar}) & \mbox{for}
\;\;\; p<0 \\[0.3cm]
0  & \mbox{for} \;\;\; p \geq 0 ;
\end{array}
\right.
\ee

\item
for ${\bf p}>0$
\be
\label{e2bb}
 W_{\bf p} (p)=
\left\{
\begin{array}{cc}
0  & \mbox{for} \;\;\; p \leq 0\\[0.3cm]
\frac{2 }{ U \hbar} J_{\frac{2{\bf p }}{U \hbar}}(\frac{2p}{ U
\hbar}) & \mbox{for} \;\;\; p>0
\end{array}
\right.
\ee

\item
and  for ${\bf p}=0$ \be \label{e2cc}
 W_{\bf 0} (p)=
\frac{1}{U \hbar} J_{0}(\frac{2p}{ U \hbar}).
\ee

\end{enumerate}

Unlike the case of the phase space with
 curvature   (\ref{ex1}), the momentum $ p $ is not a test function for
(\ref{e2aa})--(\ref{e2cc}).
 The generalized functions (\ref{e2aa})--(\ref{e2cc}) fulfill the distributed 
eigenvalue equation (\ref{31u}) although the integral
\be
\label{ee3}
\int_{-\infty}^{+\infty}p \cdot W_{\bf p}(p)dp
\ee
is not convergent.
The generalized functions (\ref{e2aa})--(\ref{e2cc}) tend to $ \delta(p-{\bf p})$ for 
$ U \rightarrow 0^+.$
Notice that we can obtain results almost immediately (\ref{e2aa})--(\ref{e2cc}) substituting 
into (\ref{e1aa})--(\ref{e1cc})
$G=iU$ or $G=-iU.$
It is natural because (\ref{ex1}) turns into (\ref{ex2}) for $G=iU$ or $G=-iU.$


\subsect{Quantization of position $ q $}

 Let us consider the problem of
eigenvalues and Wigner eigenfunctions of the position $ q $  on the
manifold $ T^*\mathbb{R} $ with a constant
 curvature of the kind (\ref{ex1}).
 The generalization of the results obtained in the case of  constant  curvature 
(\ref{ex2}) is straightforward.

Let us assume, as in the previous paragraph, that $ G $ is some positive
constant. The eigenvalue equation \be \label{.1} q*W_{\bf
q}(q,p)={\bf q}W_{\bf q}(q,p) \ee for the position $ q $ separates in
two parts: the real part \be \label{.2} q \cdot W_{\bf q}(q,p)={\bf
q}W_{\bf q}(q,p) \ee and the imaginary part \be \label{.3}
\frac{1}{2}\hbar\frac{\partial W_{\bf q}(q,p)}{\partial p} +
\frac{1}{48}G^2\hbar^3\frac{\partial^3 W_{\bf q}(q,p)}{\partial
p^3} + \cdots =0, \ee where the symbol ${\bf q}$ denotes the eigenvalue of
the position $q.$

Let us start with the equation (\ref{.3}). Multiplying it by $\frac{2}{\hbar},$
 introducing a new variable $z=\frac{p}{G \hbar}$ and defining
$ w_{\bf q}(q,z) :=\frac{\partial W_{\bf q}(q,z)}{\partial z},$
we obtain the formula
\be
\label{.4}
 w_{\bf q}(q,z) + \frac{1}{24}\frac{\partial^2 w_{\bf q}(q,z)}{\partial z^2} + \cdots =0.
\ee
This is in fact a homogeneous linear differential equation of  infinite degree. 
Its solution does not depend on the parameter $ {\bf q}$ and the factor  $G.$
The  $\lim_{G \rightarrow 0^{+}} w_{\bf q}(q,z)  $ must be $0,$ so the only one 
acceptable solution of (\ref{.4})
 is $w_{\bf q}(q,z)=0$ for every ${\bf q}.$ We see that the Wigner eigenfunction 
$ W_{\bf q}(q,z)$ depends only on $q.$
From (\ref{.2}) we immediately obtain that \be \label{.5} W_{\bf
q}(q,z)=\delta(q-{\bf q}) \qquad \forall {\bf q}. \ee

After  similar considerations for the space with
 $ 2-$ form of curvature  (\ref{ex2}), we conclude that the eigenvalues of the observable $q $ are all
of them real numbers and that their Wigner eigenfunctions are of the form (\ref{.5}).
The eigenfunctions $ W_{\bf q}(q)$ depend only on $q $ and possess  the same form as for
the flat space $\mathbb{R}^2$ with $ \Gamma=0.$


\sect{Conclusions}
\label{conc}

In the case studied by us the $*$--product of the Weyl type is a closed product.
Spectra of both observables $q,p$ are continuous and equal to
$\mathbb{R}.$ Moreover in the limit for the $2$--form of curvature
tending to $ 0^{\mp}, $ Wigner eigenfunctions go to the flat
solutions $\delta(p - {\bf p})$ and $\delta(q - {\bf q}).$

The Moyal bracket of $q$ and $p$ is \be \label{.6} \{q,p\}_{M}=1 \ee
which is the same as in the flat case and it is equal to the Poisson
bracket  $\{q , p \}_{P}.$

 It is known (see \cite{equ1,equ2}) that on every
$2$--D symplectic space all of $*$--products are
equivalent, i.e. that for every two products $*_{(1)}$ and
$*_{(2)} $ there exists such a differential operator $ \hat{T} $
that \be \label{e1} F(q,p)*_{(1)}G(q,p)=\hat{T}^{-1} \left(
\hat{T}F(q,p) *_{(2)} \hat{T}G(q,p) \right). \ee On the manifold $
T^*\mathbb{R} $ we could define $ * $--product as \be \label{e2}
F(q,p)*G(q,p) := F(q,p) \exp\left( \frac{i \hbar}{2}
\stackrel{\leftrightarrow}{\cal P} \right)G(q,p) \ee where \be
\label{e3} \stackrel{\leftrightarrow}{\cal P} :=
\frac{\stackrel{\leftarrow}{\partial}}{\partial q}
\frac{\stackrel{\rightarrow}{\partial}}{\partial p}-
\frac{\stackrel{\leftarrow}{\partial}}{\partial p}
\frac{\stackrel{\rightarrow}{\partial}}{\partial q}. \ee Such a
definition of the $*$--product is used on $ \mathbb{R}^2$ in a
system of coordinates in which the $1$--form of the connection
vanishes.

It is much easier to do any computations using the product defined by the formula (\ref{e2}) 
than those proposed by us.
 However, the explicit form of the operator $\hat{T}$ is unknown in general as there is no 
a constructive form to define it.
   In fact it is more logical to do all of calculations in
a physically motivated coordinate system although these computations may be  complicated.


\section*{Acknowledgments}
 We thank  M.
Przanowski  for interest in this work and fruitful discussions and to K. Penson, L.M. Nieto and
M. Dobrski  for their independent deductions
of formula (\ref{207q}). This work has been
partially supported by DGES of the  Ministerio de Educaci\'on y Cultura of Spain under
Project PB98-0360,  by DGI of the Ministerio de Ciencia y 
Tecnolog\'{\i}a of Spain (project BMF2002-02000 and BMF2002-03773), the Programme FEDER of the
European Community,  and the Junta de Castilla y Le\'on  (project VA085/02) (Spain). J. T.
acknowledges a la Secretar\'{\i}a del Ministerio  de Educaci\'on y Cultura of Spain for the grant
(SB2000-0129)   that supports his stay in the Universidad de Valladolid and the Batory Foundation
of Poland.  J.T. also grateful to the members of the Departamento de F\'{\i}sica  Te\'orica de la
Universidad de 
 Valladolid (Spain) for warm hospitality.


\section*{Appendix A: Proof of the formulas (\ref{313}) and (\ref{314})}

\renewcommand{\theequation}{A.\arabic{equation}}
\setcounter{equation}{0}

The $0$--form, $\sigma^{-1}(p),$ belonging to the bundle ${\cal
V}^D $ can be written as \be \label{ap1}
  \sigma^{-1}(p)= p + \frac{1}{2}p X^1 X^1 +
  \sum_{s=0}^{\infty} b_{2s+1} (X^1)^{2s} X^2,
\ee
where the coefficients $b_{2s+1}, \; s\geq 1$ are defined by the formula (\ref{309}).  For $s=0,$
 we choose $b_1=1.$
The factors $b_{2s+1}$ are real numbers
and they do not depend on the deformation parameter $\hbar.$

We know that the Wigner eigenfunction $W_{\bf p}(q,p)$ is real and differentiable
in the sense of theory of distributions. It means that  in general
\be
\label{ap2}
 \sigma^{-1}(W_{\bf p}(q,p))=\sum_{s=0}^{\infty} \sum_{a=0}^{\infty} \sum_{b=0}^{\infty}
 \hbar^{2s} w_{2s,a,b}(X^1)^a (X^2)^b.
\ee 
The coefficients $w_{2s,a,b}$ are some real distributions over the
test functions defined on  $\mathbb{R}^2.$ The real character of   $w_{2s,a,b}$ is a
straightforward consequence of (\ref{co13}). From the same
corollary, we deduce that  only even powers of $\hbar$ appear in (\ref{ap2}).

 The eigenvalue equation  for momenta (\ref{31u}) is a formal
 series in $\hbar$
 \be
 \label{ap3}
  p * W_{\bf p}(q,p)= \sum_{s=0}^{\infty} (i\hbar)^s A_{s}(q,p).
 \ee
 The functions $A_{s}(q,p)$ with $ 0 \leq s$   depend on $p, W_{\bf p}(q,p)$
 and their derivatives.
We are interested in   the explicit form of $A_{s}(q,p)$ for odd
values of $s$. From the definition of the star product, we
obviously have \be \label{ap4}
 p * W_{\bf p}(q,p)=\sigma \left( \sigma^{-1}(p)\circ \sigma^{-1}(W_{\bf
 p}(q,p))
 \right).
\ee From the expression $\sigma^{-1}(p)\circ \sigma^{-1}(W_{\bf
p}(q,p))$ only
 elements not
containing $X$'s give a contribution to $A_{s}(q,p). $
 It means that an element from the series
$\sigma^{-1}(p)$ containing $(X^1)^a (X^2)^b$   gives a
contribution to  $ p * W_{\bf p}(q,p)$  only if   it is multiplied
by  $w_{2s,a,b}(X^1)^b (X^2)^a.$    Moreover, to obtain the
functions $A_{2s+1}(q,p) $ with odd index $2s+1$ the additional
condition  $b+a=2k+1, \; k \leq s $ must be fulfilled. Looking at
(\ref{310}), we see that
\be\label{ap4.5}
\begin{array}{l}
(i \hbar)^{2s+1} A_{2s+1}(q,p) = \sigma \left( b_{2s+1}(X^1)^{2s}X^2
\circ w_{0,1,2s} X^1(X^2)^{2s} \right.
\\[0.3cm]
\qquad \left. +\; b_{2s-1}(X^1)^{2s-2}X^2 \circ  \hbar^2
w_{2,1,2s-2} X^1(X^2)^{2s-2} + \cdots  + X^2 \circ
\hbar^{2s}w_{2s,1,0}X^1 \right) . 
\end{array}\ee 
Let us analyze more
carefully the term $\sigma \left(b_{2p+1}(X^1)^{2p}X^2\circ
w_{2k,1,2p} X^1(X^2)^{2p}\right).$ From  the definition (\ref{13})
of  $\circ$--product we see that 
\be \label{ap5}
  \sigma \left(b_{2p+1}(X^1)^{2p}X^2\circ w_{2k,1,2p}
  X^1(X^2)^{2p}\right)=  b_{2p+1} w_{2k,1,2p}
  (\omega^{12})^{2p}\omega^{21}\left(\frac{i\hbar}{2}\right)^{2p+1}
  \frac{(2p)!(2p)!}{(2p+1)!}.
\ee
 Using     (\ref{Poisson})  we obtain that formula (\ref{ap5})
 equals
 \be
 \label{ap6}
(-1) b_{2p+1} w_{2k,1,2p} \left(\frac{i\hbar}{2}\right)^{2p+1} \frac{(2p)!}{2p+1}.
 \ee
 We see that the right side of the equation   (\ref{ap4.5})  is
 the sum
 \be
 \label{ap7}
 \sum_{p=0}^{s}  (-1) b_{2p+1} w_{2s-2p,1,2p} \hbar^{2s-2p} \left(\frac{i\hbar}{2}\right)^{2p+1} 
\frac{(2p)!}{2p+1}.
 \ee
 Comparing (\ref{ap4.5}) and (\ref{ap7}) we see that
 \bea \label{ap8}
   A_{2s+1}(q,p) &=& \sum_{p=0}^{s} (-1) b_{2p+1} w_{2s-2p,1,2p}  \:
   \frac{1}{i^{2s-2p}} \:\frac{1}{2^{2p+1}}\: \frac{(2p)!}{2p+1} \nonumber\\[0.3cm]
& =&\sum_{p=0}^{s}(-1)^{s-p+1}  b_{2p+1} w_{2s-2p,1,2p} \:\frac{1}{2^{2p+1}}\:
   \frac{(2p)!}{2p+1}\nonumber\\[0.3cm]
&=&\frac{(-1)^{s+1}}{2} \left(  w_{2s,1,0} +
   \sum_{p=1}^{s}(-1)^{-p}  b_{2p+1} w_{2s-2p,1,2p} \:\frac{1}{2^{2p}}\:
   \frac{(2p)!}{2p+1}
\right).
\eea

Let us consider the element $\hbar^{2s}w_{2s,1,0}X^1.$    From the
recurrent relation (\ref{34}) using the definition (\ref{delta-1})
of the operator $\delta^{-1},$ we deduce that only three kinds of
terms may generate this element.
\begin{itemize}
\item
The  derivative $\frac{\partial w_{2s,0,0} }{\partial q}.$ Here we
should note that this term is different from $0$ only for $s=0.$
Indeed, the elements $w_{2s,0,0}, \; s > 0$, can be obtained only from
terms on which the operator $\delta^{-1}$ acts. This operator
generates $X$'s.
\item
The element $w_{2s,1,0}$ itself. Note that the following
commutator gives \be \label{ap9}
 [X^2 dq,w_{2s,1,0}X^1]=2 \omega^{21}\left(\frac{i
\hbar}{2}\right)w_{2s,1,0}dq \stackrel{\rm (\ref{Poisson})}{=}  - (i \hbar)  w_{2s,1,0}dq.
\ee
\item
The expressions $w_{2s-2p,1,2p}, \; s \geq p \geq 1 $, through
their commutators with $r$ (\ref{33}). Indeed in the commutator
$[r,w_{2s-2p,1,2p}]$  the following element also  appears \[ 2
(-1)^{p+1}\: a_{2p+1}\:
w_{2s-2p,1,2p}\:(\omega^{12})^{2p}\:\omega^{21}\left(\frac{i
\hbar}{2}\right)^{2p+1}\frac{1}{(2p+1)!}(2p)! (2p)! dq,
\]
and using (\ref{Poisson})  and (\ref{309}) we can write the above
expression  as \be \label{ap10}
 -
b_{2p+1}\: w_{2s-2p,1,2p}\: \frac{1}{2^{2p}}\:(i \hbar)^{2p+1}\:  \frac{(2p)!}{2p+1}dq.
\ee
\end{itemize}
Taking together (\ref{ap9}) and (\ref{ap10})   we see that   for $ s \geq 1 $
\[
 \hbar^{2s}w_{2s,1,0}X^1 = \frac{\delta^{-1}}{i
 \hbar}\left(
 - i \hbar \:\hbar^{2s}\: w_{2s,1,0}dq - \sum_{p=1}^{s}
b_{2p+1}\: w_{2s-2p,1,2p}\: \frac{1}{2^{2p}}\:(i \hbar)^{2p+1}\:
\frac{(2p)!}{2p+1}dq
 \right).
\]
So, finally \be \label{ap11} w_{2s,1,0}X^1=- \sum_{p=1}^{s} (-1)^p
\:b_{2p+1} \: w_{2s-2p,1,2p} \:     \frac{1}{2^{2p}}
 \frac{(2p)!}{2p+1}.
\ee
Putting (\ref{ap11}) into   (\ref{ap8})  we obtain that
\be
\label{ap12}
A_{2s+1}(q,p)= \left\{ \begin{array}{ll}
-\frac{1}{2} \: w_{0,1,0}  & {\rm if \;\; s=0,} \\[0.3cm]
0  & {\rm if \;\; s>0.}
\end{array}
\right.
\ee


\section*{Appendix B: Proof of the relation
(\ref{317})}

\renewcommand{\theequation}{B.\arabic{equation}}

\setcounter{equation}{0}

Let us start our considerations with the analysis of the recurrent
formula \be \label{ac1} c_m=\sum_{i=1}^{m-1}C(i,m-i) c_i c_{m-i}
\ee where the coefficients $C(i,m-i)$  fulfill the following
relation: 
\be \label{ac3} 
C(i,m-i)+ C(m-i,i)=1,   \qquad  1\leq i
\leq m-1. 
\ee 
It is easy to note that the coefficients $a_{2m+1}$
defined by (\ref{207})
 with the initial condition $a_3=\frac{1}{8}$ are of the kind
 (\ref{ac1}). We need only to write
 \[
 c_m = a_{2m+1} \qquad{\rm for } \qquad m\geq 1.
 \]
 Due to the fact that the equation (\ref{ac3}) holds, each formula
 of the form (\ref{ac1}) is equivalent    to the following one
 \be
  \label{ac4}
  c_m= \frac{1}{2} \sum_{i=1}^{m-1} c_i c_{m-i}.
  \ee

\vspace{0.5cm}

The eigenvalue equation for the  momentum $p$ (\ref{315}) may be
written as
 \be
  \label{ac5}
  - p \cdot L W_{\bf p}(p)= {\bf p}W_{\bf p}(p),
  \ee
  where $L$ denotes the differential operator
  \be
  \label{ac6}
  L =\left(\sum_{i=0}^{\infty}a_{2i+1}\hbar^i \frac{d^{2i}}{d p^{2i}}
  \right).
  \ee
  The coefficients $a_{2i+1}$ for $i \geq 1$ are defined by the
  formula (\ref{207q}). For $i=0$ we put $a_1=-1.$

  The eigenvalue equation for $p^2=p*p$ is
  \be
  \label{ac7}
     p \cdot L\left( p \cdot L W_{\bf p}(p) \right) = {\bf p}^2 W_{\bf
     p}(p).
  \ee
  The relation holds
  \be
  \label{ac8}
    \frac{d^{k}p U(p)}{d p^{k}}= p \frac{d^{k} U(p)}{d p^{k}} + k \frac{d^{k-1} U(p)}{d p^{k-1}}
    , \qquad k \in {\mathbb N}
  \ee
  where $U(p)$ denotes a differentiable function of $p$.
Applying (\ref{ac6}) and (\ref{ac8})  we see that
\be\label{ac9}
\begin{array}{r}
  p \cdot L\left( p \cdot L W_{\bf p}(p) \right)=   p^2 \displaystyle
\sum_{m=0}^{\infty} \sum_{k=0}^{\infty}
  a_{2m+1}a_{2k+1} \hbar^{2m+2k }  \frac{d^{2m+2k} W(p)}{d p^{2m+2k}}\;
  \\[0.4cm]
 +\; 2p \displaystyle
\sum_{m=0}^{\infty} \sum_{k=0}^{\infty}\: m \:
a_{2m+1}a_{2k+1} \hbar^{2m+2k }  \frac{d^{2m+2k-1} W(p)}{d
p^{2m+2k-1}}, 
\end{array}\ee 
where the coefficient  of $\hbar^{2n }
\frac{d^{2n} W(p)}{d p^{2n}}$ is
 \be
 \label{ac10}
     s_{2n}= p^2 \sum_{m=0}^{n} a_{2m+1}a_{2n-2m+1}.
 \ee
 In the case $n>0$ we can write (\ref{ac10}) as
 \be
 \label{ac105}
s_{2n}=p^2\left( \sum_{m=1}^{n-1}
     a_{2m+1}a_{2n-2m+1}+ 2a_{2n+1}a_1 \right).
 \ee
 The factors $a_{2m+1}$  fulfill the recurrent relation (\ref{ac4})
 and $a_1=-1.$
 Hence, we obtain that
 \be
 \label{ac11}
   s_{2n}= \left\{ \begin{array}{ll}
   p^2 & {\rm for} \;\; n=0,  \\[0.3cm]
   - \frac{1}{4}p^2 & {\rm for} \;\; n=1, \\[0.3cm]
   0 & {\rm for} \;\; n>1.
   \end{array}
   \right.
 \ee
By $r_{2n}$ we will denote the coefficients appearing in
(\ref{ac9}) associated with $\hbar^{2n }  \frac{d^{2n-1} W(p)}{d
p^{2n-1}}$ 
\be \label{ac12}
 r_{2n}=2p \sum_{m=0}^{n}   \: m \: a_{2m+1}a_{2n-2m+1}.
\ee
For $n>0$ we can write
\be
\label{ac13}
 r_{2n}=2p \left( \sum_{m=1}^{n-1}   \: m \: a_{2m+1}a_{2n-2m+1} + n a_{2n+1}a_1 \right).
\ee Again the sum on the right side  is an expression of the type
(\ref{ac4})  $\frac{m+(n-m)}{n}=1$. So, we can see that
 \be
 \label{ac14}
   r_{2n}= \left\{ \begin{array}{ll}
   0 & {\rm for} \;\; n=0,  \\[0.3cm]
   - \frac{1}{4}p & {\rm for} \;\; n=1, \\[0.3cm]
   0 & {\rm for} \;\; n>1.
   \end{array}
   \right.
 \ee
 Taking together (\ref{ac11})  and (\ref{ac14}) we finally obtain that
 \be
 \label{ac15}
  p \cdot L\left( p \cdot L W_{\bf p}(p) \right)=p^2 W_{\bf p}(p) -
  \frac{1}{4} \: p \:\hbar^{2 }  \frac{d W(p)}{d
p} - \frac{1}{4} \: p^2\: \hbar^{2}  \frac{d^{2} W(p)}{d p^{2}}.
 \ee


\section*{Appendix C: The flat limit of  Wigner eigenfunctions for  momentum ${\bf p}$}

\renewcommand{\theequation}{C.\arabic{equation}}

\setcounter{equation}{0}

Let us consider first  the phase space with a constant
 curvature tensor like (\ref{ex1}). For the eigenvalue $\;{\bf p}>0, \; $ the Wigner 
eigenfunction $W_{\bf p}(p)$ is
defined by the formula (\ref{e1bb}). The Fourier transform of
$W_{\bf p}(p)$ equals (see (\ref{??2})) \be \label{pB1} {\cal
F}[W_{\bf p}](\lambda)=\int_0^{\infty}
   \frac{4 }{G \pi  \hbar} \cosh{\left(\frac{{\bf p} \pi }{ G \hbar}\right)}
    K_{\frac{2i{\bf p }}{G \hbar}}\left(\frac{2p}{ G \hbar}\right) e^{-2 \pi i \lambda
    p}d p.
\ee Using the Mathematica programme we can see that \be
\label{pB2}
 {\cal F}[W_{\bf p}](\lambda)=  \frac{1}{\sinh{\left(\frac{{\bf p} \pi }{ G \hbar}\right)}
 \sqrt{1+G^2\:\hbar^2\:\pi^2\:\lambda^2}}
 \sinh{\left(\frac{{\bf p}\pi}{G \hbar}-\frac{2i{\bf p}}{G \hbar}{\rm arcsinh}(\pi G \hbar \lambda)
 \right)}.
 \ee
 Let us compute the limit  of the function  ${\cal F}[W_{\bf
 p}](\lambda)$ for $G \rightarrow 0^+ $ in the distributional sense.
 Take the test function $\varphi(\lambda) \in {\cal S}(\lambda).$
The relation holds
\be\begin{array}{l}\label{pB3}
\sinh{\left(\frac{{\bf p}\pi}{G \hbar}-\frac{2i{\bf p}}{G
\hbar}{\rm arcsinh}(\pi G \hbar \lambda)
 \right)}\\[0.3cm]
\;\; =\frac{1}{2}\left(  \exp{(\frac{{\bf
p}\pi}{G \hbar})}\exp{(-\frac{2i{\bf p}}{G \hbar}{\rm arcsinh}(\pi
G \hbar \lambda))} - \exp{(-\frac{{\bf p}\pi}{G
\hbar})}\exp{(\frac{2i{\bf p}}{G \hbar}{\rm arcsinh}(\pi G \hbar
\lambda))}
 \right).
\end{array}\ee 
For any real ${\bf p},G,\hbar, \lambda$ the following
inequalities are true 
\be
 \label{pB4}
  \left|\exp{(-\frac{2i{\bf p}}{G \hbar}{\rm
arcsinh}(\pi G \hbar \lambda))} \right|\leq 1 ,\qquad
\left|\exp{(\frac{2i{\bf p}}{G \hbar}{\rm arcsinh}(\pi G \hbar
\lambda))} \right|\leq 1.
 \ee
Then, \be \label{pB5} \langle\langle {\cal F}[W_{\bf p}](\lambda),
\varphi(\lambda) \rangle\rangle=\int_{-\infty}^{+\infty}{\cal
F}[W_{\bf p}](\lambda) \varphi(\lambda) d \lambda. 
\ee 
Using
({\ref{pB3}), we can write
\be\begin{array}{r}\label{pB6}
\displaystyle
  \int_{-\infty}^{+\infty}{\cal F}[W_{\bf
p}](\lambda) \varphi(\lambda) d \lambda=
\int_{-\infty}^{+\infty}\frac{\exp{(\frac{{\bf p}\pi}{G
\hbar})}\exp{(-\frac{2i{\bf p}}{G \hbar}{\rm arcsinh}(\pi G \hbar
\lambda))}}{2\sinh{\left(\frac{{\bf p} \pi }{ G \hbar}\right)}
 \sqrt{1+G^2\:\hbar^2\:\pi^2\:\lambda^2}}               \varphi(\lambda) d \lambda
\\[0.5cm]
 +\displaystyle \int_{-\infty}^{+\infty}\frac{\exp{(-\frac{{\bf p}\pi}{G
\hbar})}\exp{(\frac{2i{\bf p}}{G \hbar}{\rm arcsinh}(\pi G \hbar
\lambda))}}{2\sinh{\left(\frac{{\bf p} \pi }{ G \hbar}\right)}
 \sqrt{1+G^2\:\hbar^2\:\pi^2\:\lambda^2}}               \varphi(\lambda) d
 \lambda .
\end{array}\ee 
We have the estimation   (see (\ref{pB4}))
 \be
 \label{pB7}
\left| \int_{-\infty}^{+\infty}\frac{ \exp{(\frac{2i{\bf p}}{G
\hbar}{\rm arcsinh}(\pi G \hbar \lambda))}}{
 \sqrt{1+G^2\:\hbar^2\:\pi^2\:\lambda^2}}               \varphi(\lambda) d
 \lambda \right| <  \int_{-\infty}^{+\infty} |\varphi(\lambda)   |
 d \lambda.
\ee 
The function $\varphi(\lambda)$ belongs to the
Schwartz space, so the integral
$\int_{-\infty}^{+\infty}|\varphi(\lambda)   |
 d \lambda$ is convergent. Let it be equal to $A.$
 It means that the absolute value of the second term of the integral (\ref{pB6})
 is not bigger than
 $
\frac{\exp{(-\frac{{\bf p}\pi}{G \hbar})}}{2\sinh{\left(\frac{{\bf
p} \pi }{ G \hbar}\right)}}A.
 $
But \be \label{pB8} \lim_{G \rightarrow
0^+}\frac{\exp{(-\frac{{\bf p}\pi}{G
\hbar})}}{2\sinh{\left(\frac{{\bf p} \pi }{ G \hbar}\right)}}A =0,
\ee 
hence, in the limit $G \rightarrow
                  0^+ $
only the first integral in (\ref{pB6}) can be different from $0$.

Let us analyze the expression \be \label{pB9}
  \int_{-\infty}^{+\infty}\frac{
\exp{(-\frac{2i{\bf p}}{G \hbar}{\rm arcsinh}(\pi G \hbar
\lambda))}}{
 \sqrt{1+G^2\:\hbar^2\:\pi^2\:\lambda^2}}               \varphi(\lambda) d
 \lambda,
\ee appearing in the first part of $\langle\langle {\cal F}[W_{\bf
p}](\lambda), \varphi(\lambda) \rangle\rangle$ (\ref{pB5}).
Introducing the new variable defined by $z = \frac{{\rm
arcsinh}(\pi G \hbar \lambda)}{G \pi \hbar } $ and remembering
that $\frac{d \: {\rm arcsinh}(x)}{dx}= \frac{1}{\sqrt{1 +x^2}}$
we obtain that (\ref{pB9}) is \be \label{pB10}
  \int_{-\infty}^{+\infty} \exp{(-2 i \pi {\bf p}z)} \varphi\left(\frac{\sinh{(\pi G \hbar z)}}{\pi G \hbar}
\right)
  dz.
\ee The following property holds \be \label{pB11} \left|
\frac{\sinh{(\pi G \hbar z)}}{\pi G \hbar} \right| \geq  |z|. \ee
The function $\varphi(z)$ belongs to the Schwartz space, hence for
every $k \in {\mathbb N}$ it is possible to find a real number $z_0$
so that for every $z$ fulfilling the inequality $|z| > z_0$ the
relation \be \label{pB12}
    |\varphi(z)| < \left| \frac{1}{z^k} \right|
\ee holds. According to (\ref{pB11}),  if  $|z| > z_0,$ it is also
true that 
\be \label{pB13}
    \left|\varphi\left(\frac{\sinh{(\pi G \hbar z)}}{\pi G \hbar}\right)\right| < \left| \frac{1}{z^k}
    \right|.
\ee 
Let $k=2.$ The function \be \label{pB14} F(z):=\left\{
\begin{array}{ll}
\max_{(-\infty,+\infty)}{|\varphi|} & \; {\rm for} |z|<z_0, \\[0.3cm]
  \frac{1}{z^2}  & \; {\rm for} |z| > z_0
\end{array}
\right.
\ee
 is integrable and for all $G>0$
 \be
 \label{pB15}
   \left|\varphi\left(\frac{\sinh{(\pi G \hbar z)}}{\pi G
   \hbar}\right)\right|\leq F(z)
 \ee
 and the Lebesgue theorem (\cite{asp})  implies that
\be\begin{array}{l}\label{pB17} 
 \lim_{G \rightarrow 0^+} \left[\displaystyle
 \int_{-\infty}^{+\infty} \exp{(-2 i \pi {\bf p}z)} \varphi
\left(\frac{\sinh{(\pi G \hbar z)}}{\pi G \hbar}\right)
  dz  \right] 
\\[0.4cm]
\qquad = \displaystyle 
\int_{-\infty}^{+\infty}  \left[ \lim_{G \rightarrow
0^+}  \exp{(-2 i \pi {\bf p}z)}
   \varphi\left(\frac{\sinh{(\pi G \hbar z)}}{\pi G \hbar}  \right)
   \right]
   dz.
 \end{array}\ee
 Moreover,
   \be
\label{pB18}
\lim_{G \rightarrow 0^+} \frac{\sinh{(\pi G \hbar z)}}{\pi G \hbar}=z
 \ee
 and, finally, we can write  the limit of the integral (\ref{pB10})
 when $G \rightarrow 0^+$ as
 \be
 \label{pB19}
    \int_{-\infty}^{+\infty}  \exp{(-2 i \pi {\bf p}z)}
    \varphi(z)dz.
 \ee
Using the relation \be \label{pB20}
 \lim_{G \rightarrow 0^+} \frac{\exp{\left(\frac{{\bf p}\pi}{G
 \hbar}\right)}}
 {2 \sinh{\left(\frac{{\bf p}\pi}{G \hbar}\right)}} =1
 \ee
 from (\ref{pB6})  according to (\ref{pB8}), (\ref{pB19}) and
 (\ref{pB20}), we conclude that
 \be
 \label{pB21}
  \lim_{G \rightarrow 0^+} \int_{-\infty}^{+\infty}{\cal F}[W_{\bf
p}](\lambda) \varphi(\lambda) d \lambda = \int_{-\infty}^{+\infty}
\exp{(-2 i \pi {\bf p} \lambda)}
    \varphi(\lambda)d\lambda.
 \ee
 It means that
 \be
 \label{pB210}
\lim_{G \rightarrow 0^+}{\cal F}[W_{\bf p}](\lambda)= \exp{(-2 i
\pi {\bf p} \lambda)}.
 \ee
 We recall that
 \be
 \label{pB22}
 \exp{(-2 i
\pi {\bf p} \lambda)}= {\cal F}[\delta(p-{\bf p})], 
\ee 
hence, the
flat limit of the Wigner eigenfunction $W_{\bf p}(p)$ of momenta
$p$ for positive eigenvalue ${\bf p}$ is just $\delta(p-{\bf p}).$

\vspace{0.5cm} The proof for the case ${\bf p}<0$ is analogous.
\vspace{0.5cm}

In the situation in which ${\bf p}=0,$ the Fourier transform is 
\be
\label{pB23}
   {\cal F}[W_{\bf 0}](\lambda)= \frac{1}{\sqrt{1+
   G^2\:\hbar^2\:\pi^2\:\lambda^2}},
\ee 
hence 
\be \label{pB24} 
\langle\langle {\cal F}[W_{\bf
0}](\lambda), \varphi(\lambda) \rangle\rangle
=\int_{-\infty}^{+\infty}\frac{1}{\sqrt{1+
   G^2\:\hbar^2\:\pi^2\:\lambda^2}}
 \varphi(\lambda) d \lambda.
\ee 
Since  the function $\varphi(\lambda) \in {\cal S(\lambda)},$ for
every $k \in {\mathbb N}$ it is possible to find such $\lambda_0$
that for every $\lambda$ fulfilling the inequality $|\lambda| >
\lambda_0$ the relation (\ref{pB12})   holds.  Let $k=2.$ The
function $F(\lambda)$ defined as (\ref{pB14}) is integrable and
for all $G>0$
 \be
 \label{pB25}
   \left|\frac{1}{\sqrt{1+
   G^2\:\hbar^2\:\pi^2\:\lambda^2}}
 \varphi(\lambda)
\right|\leq F(\lambda).
 \ee
 This gives with the Lebesgue theorem that
 \[
 \lim_{G \rightarrow 0^+}\left[
\int_{-\infty}^{+\infty}\frac{1}{\sqrt{1+
   G^2\:\hbar^2\:\pi^2\:\lambda^2}}
 \varphi(\lambda) d \lambda \right] = \int_{-\infty}^{+\infty}
\left[\lim_{G \rightarrow 0^+}\frac{1}{\sqrt{1+
   G^2\:\hbar^2\:\pi^2\:\lambda^2}}
 \varphi(\lambda)\right] d \lambda,
\] 
which means that 
\be
 \label{pB27}
\lim_{G \rightarrow 0^+}{\cal F}[W_{\bf 0}](\lambda)=1 
\ee 
and,
hence 
\be
 \label{pB28}
\lim_{G \rightarrow 0^+}W_{\bf 0}(p)= \delta(p).
 \ee

The case in which the curvature  is given by (\ref{ex2}) is more
difficult for rigorous mathematical analysis. Hence,  we have
decided to present an intuitive proof that indicates that Wigner
eigenfunctions (\ref{e2aa})--(\ref{e2cc}) should tend to Dirac
deltas when $U$ goes to $0^+$.

As we can easily check   for $p>1,$ fast oscillations appear and
their frequency  increases if $U$ tends to $0^+. $ It means that
for $U$'s close to $0$ the integrals
\[
\int_{1}^{+\infty}\frac{2}{U\hbar}J_\frac{2}{u \hbar
}\left(\frac{2p}{U \hbar}\right)\varphi(p)dp \]
 are arbitrarily small for each
test function $\varphi(p).$ Moreover, for   $p<1,$ the function
$\frac{2}{U\hbar}J_\frac{2}{u \hbar }\left(\frac{2p}{U
\hbar}\right)$ goes to $0$ as $U \rightarrow 0^+.$

Taking into account these properties of the function
$\frac{2}{U\hbar}J_\frac{2}{u \hbar }\left(\frac{2p}{U
\hbar}\right)$ we arrive to the conclusion that for $U$'s close to
$0$ the integrals 
\be\label{ad1}
 \int_{0}^{\infty}\frac{2}{U\hbar}J_\frac{2}{u
\hbar }\left(\frac{2p}{U \hbar}\right)\varphi(p)dp\; \simeq\;
\int_{1-\epsilon}^{1+\epsilon}\frac{2}{U\hbar}J_\frac{2}{u \hbar
}\left(\frac{2p}{U \hbar}\right)\varphi(p) d p,
\ee
 where $\epsilon$ is
some real positive number.
 Let us consider  the difference 
\be
\label{ico} \left| \int_{0}^{\infty}\frac{2}{U\hbar}J_\frac{2}{u
\hbar }\left(\frac{2p}{U \hbar}\right)\varphi(p)dp  -
\int_{1-\epsilon}^{1+\epsilon}\frac{2}{U\hbar}J_\frac{2}{u \hbar
}\left(\frac{2p}{U \hbar}\right)\varphi(p) d p \: \right| < A, 
\ee
where $A$ is any fixed positive real number. Let us solve the
inequality (\ref{ico}) for two  fixed positive numbers $U_1, U_2$
such that $ U_1 < U_2. $ The  minimal $\epsilon_1$ fulfilling
(\ref{ico}) for $U_1$ is smaller than $\epsilon_2$ being the
minimal number for which (\ref{ico}) holds in case $U=U_2.$

For each fixed $\epsilon$ integrals 
\be \label{ad2}
\int_{1-\epsilon}^{1+\epsilon}\frac{2}{U\hbar}J_\frac{2}{u \hbar
}\left(\frac{2p}{U \hbar}\right) d p 
\ee 
go to $1$ as $U
\rightarrow 0$. So, we can say that in fact the flat limit of
(\ref{e2cc}) is the Dirac delta $\delta(p-{\bf p})$.

Similar considerations can be done for ${\bf p}<0$ and ${\bf
p}=0.$


\end{document}